\documentclass[fleqn,usenatbib]{rasti}

\usepackage{newtxtext,newtxmath,siunitx, multirow}
\usepackage{lineno}
\usepackage{graphicx}
\usepackage{subcaption}

\usepackage[T1]{fontenc}

\DeclareRobustCommand{\VAN}[3]{#2}
\let\VANthebibliography\thebibliography
\def\thebibliography{\DeclareRobustCommand{\VAN}[3]{##3}\VANthebibliography}

\usepackage{graphicx}	
\usepackage{amsmath}	
\usepackage{acro} 
\acsetup{
  uppercase/list
}
\usepackage{longtable}
\usepackage{booktabs}
\usepackage{ragged2e}
\newcolumntype{L}[1]{>{\raggedright\arraybackslash}p{#1}}
\usepackage{gensymb}
\usepackage{orcidlink}
\usepackage{hyperref}

\DeclareAcronym{nuv}{
  short=NUV,
  long=near-ultraviolet,
}
\DeclareAcronym{uv}{
  short=UV,
  long=ultraviolet,
}
\DeclareAcronym{iue}{
  short=IUE,
  long=International Ultraviolet Explorer,
}
\DeclareAcronym{hst}{
  short=HST,
  long=Hubble Space Telescope,
}
\DeclareAcronym{cbe}{
  short=CBe,
  long=classical Be,
}
\DeclareAcronym{ir}{
  short=IR,
  long=infrared,
}
\DeclareAcronym{nrp}{
  short=NRP,
  long=non-radial pulsation,
}
\DeclareAcronym{sed}{
  short=SED,
  long=spectral energy distribution,
}
\DeclareAcronym{snr}{
  short=SNR,
  long=signal-to-noise ratio,
}
\DeclareAcronym{cae}{
  short=CAe,
  long=classical Ae,
}
\DeclareAcronym{cme}{
  short=CME,
  long=coronal mass ejection,
}
\DeclareAcronym{tts}{
  short=TTS,
  long=T Tauri star,
}
\DeclareAcronym{ctts}{
  short=CTTS,
  long= classical T Tauri star,
}
\DeclareAcronym{hwo}{
  short=HWO,
  long=Habitable Worlds Observatory,
}
\DeclareAcronym{fuv}{
  short=FUV,
  long=far-ultraviolet,
}
\DeclareAcronym{euv}{
  short=EUV,
  long=extreme ultraviolet,
}
\DeclareAcronym{nir}{
  short=NIR,
  long=near-infrared,
}
\DeclareAcronym{bss}{
  short=BSS,
  long=blue straggler star,
}
\DeclareAcronym{yss}{
  short=YSS,
  long=yellow straggler star,
}
\DeclareAcronym{rss}{
  short=RSS,
  long=red straggler star,
}
\DeclareAcronym{ssg}{
  short=SSG,
  long=sub-sub giant,
}
\DeclareAcronym{bssl}{
  short=BSSL,
  long=Blue Skies Space Ltd.,
}
\DeclareAcronym{hpic}{
  short=HPIC,
  long=Preliminary Input Catalog,
}

\title[The Preliminary Mauve Science Programme]{The Preliminary Mauve Science Programme: Science themes identified for the first year of operations}

\author[Mauve Science Collaboration - Year 1]{Mauve Science Collaboration - Year 1:
Marcel A. Agüeros$^{1}$ \orcidlink{0000-0001-7077-3664},
Don Dixon$^{2}$ \orcidlink{0000-0001-6977-9495},
Chuanfei Dong$^{3,4}$ \orcidlink{0000-0002-8990-094X},
\newauthor
Girish M. Duvvuri$^{2}$ \orcidlink{0000-0002-7119-2543},
Patrick F. Flanagan$^{5}$ \orcidlink{0009-0001-6677-5090},
Christopher M. Johns–Krull$^{6}$ \orcidlink{0000-0002-8828-6386},
Hongpeng Lu$^{3}$ \orcidlink{0009-0002-9364-8844},
\newauthor
Hiroyuki Maehara$^{7, 8, 9}$ \orcidlink{0000-0003-0332-0811},
Kosuke Namekata$^{7,8,10,11}$ \orcidlink{0000-0002-1297-9485},
Alejandro Núñez$^{1}$ \orcidlink{0000-0002-8047-1982},
Elena Pancino$^{12}$ \orcidlink{0000-0003-0788-5879},
\newauthor
Sharmila Rani$^{12}$ \orcidlink{0000-0003-4233-3180},
Anusha Ravikumar$^{13}$ \orcidlink{0000-0002-9015-6417},
T.\ A.\ A.\ Sigut$^{13}$ \orcidlink{0000-0002-0803-8615},
Keivan G. Stassun$^{2}$ \orcidlink{0000-0002-3481-9052},
Jamie J. Stewart$^{5}$ \orcidlink{0009-0002-7515-6213},
\newauthor
Krisztián Vida$^{14}$ \orcidlink{0000-0002-6471-8607},
Emma T. Whelan$^{5}$ \orcidlink{0000-0002-3741-9353},
Benjamin J. Wilcock$^{15}$ \orcidlink{0000-0002-4530-3676},
Sharafina Razin$^{15}$ \orcidlink{0009-0001-9170-8690},
Arianna Saba$^{15,16}$ \orcidlink{0000-0002-1437-4228},
\newauthor
Ian Stotesbury$^{15}$ \orcidlink{0000-0002-4902-5271},
Giovanna Tinetti$^{15,16}$ \orcidlink{0000-0001-6058-6654},
Marcell Tessenyi$^{15}$ \orcidlink{0000-0002-3905-5944},
Jonathan Tennyson$^{17,15}$ \orcidlink{0000-0002-4994-5238}\\
\\
$^{1}$Department of Astronomy, Columbia University, 550 West 120th Street, New York, NY 10027.\\
$^{2}$Department of Physics and Astronomy, Vanderbilt University, Nashville, TN 37235, USA\\
$^{3}$Department of Astronomy, Boston University, Boston, MA 02215, USA \\
$^{4}$School of Natural Sciences, Institute for Advanced Study, Princeton, NJ 08540, USA \\
$^{5}$Department of Physics, Maynooth University, Maynooth, Co Kildare, Ireland\\
$^{6}$Rice University, Department of Physics and Astronomy, Brockman Hall for Physics Suite 201, 6100 Main Street, Houston, Texas 77005, USA\\
$^{7}$The Hakubi Center for Advanced Research, Kyoto University, Yoshida-Honmachi, Sakyo-ku, Kyoto 606-8501, Japan \\
$^{8}$Department of Physics, Kyoto University, Kitashirakawa-Oiwake-cho, Sakyo-ku, Kyoto, 606-8502, Japan \\
$^{9}$Subaru Telescope Okayama Branch Office, National Astronomical Observatory of Japan, National Institutes of Natural Sciences \\
$^{10}$Heliophysics Science Division, NASA Goddard Space Flight Center, 8800 Greenbelt Road, Greenbelt, MD 20771, USA \\
$^{11}$The Catholic University of America, 620 Michigan Avenue, N.E. Washington, DC 20064, USA \\
$^{12}$INAF - Osservatorio Astrofisico di Arcetri, Largo Enrico Fermi 5, I-50125 Firenze, Italy\\
$^{13}$Department of Physics $\&$ Astronomy, Institute for Earth $\&$ Space Exploration, The University of Western Ontario, \\ 1151 Richmond Street, London, Ontario, N6A 3K7 Canada \\
$^{14}$Konkoly Observatory, Konkoly-Thege Miklós út 15-17, Budapest, 1121 Hungary\\
$^{15}$Blue Skies Space Ltd., 69 Wilson Street, London, EC2A 2BB UK \\
$^{16}$Department of Physics, King’s College London, Strand, London WC2R 2LS, UK \\
$^{17}$Department of Physics and Astronomy, University College London, Gower Street, WC1E 6BT London, UK\\
}

\date{Accepted XXX. Received YYY; in original form ZZZ}

\pubyear{\the\year{}}

\begin{document}
\label{firstpage}
\pagerange{\pageref{firstpage}--\pageref{lastpage}}
\maketitle

\begin{abstract}
Mauve is a low-cost small satellite developed and operated by Blue Skies Space Ltd. The payload features a 13 cm telescope connected with a fibre that feeds into a UV-Vis spectrometer. The detector covers the 200-700 nm range in a single shot, obtaining low resolution spectra at \textit{R}$\sim$20-65. Mauve has launched on 28th November 2025, reaching a 510 km Low-Earth Sun-synchronous orbit. The satellite will enable UV and visible observations of a variety of stellar objects in our Galaxy, filling the gaps in the ultraviolet space-based data. The researchers that have already joined the mission have defined the science themes, observational strategy and targets that Mauve will observe in the first year of operations. To date 10 science themes have been developed by the Mauve science collaboration for year 1, with observational strategies that include both long duration monitoring and short cadence snapshots. Here, we describe these themes and the science that Mauve will undertake in its first year of operations. 

\end{abstract}




\section{Introduction}

The study of the universe through ultraviolet (UV) radiation has enabled many breakthroughs in the past decades.
Space telescopes with UV capabilities, such as the \ac{iue} \citep{1978Natur.275..372B}, the \ac{hst} and UVIT@Astrosat \citep{2012SPIE.8443E..1NK}, have provided decades of valuable astrophysical data. However, as these facilities have either been decommissioned or are highly oversubscribed,  the UV region of the electromagnetic spectrum remains under-sampled, with existing datasets often fragmentary and infrequently acquired.

To address this gap, \ac{bssl} has developed Mauve, a small satellite, using high-heritage, off-the-shelf technology that covers a broad spectral range, simultaneously spanning the \ac{nuv} and visible wavelengths. 

Pioneering a new model for science satellites and space science  \citep{archer2020sustainable}, Mauve launched on 28th November 2025 and will begin a collaborative scientific programme in early 2026, that focuses on monitoring stellar activity and variability. 
A major strength of the Mauve science programme is the availability of thousands of observational hours each year for its consortium members. {This makes it ideal for time-domain astronomy, where hundreds of hours can be dedicated to both, long-duration observations and repeat observations}. Additionally, Mauve will be ideally placed to conduct pilot studies and high-risk investigations. 

Providing constraints on stellar activity, variability, and the influence a host star has on its local environment and potential planet habitability are a few examples of the areas Mauve will contribute to over its three-year lifetime. 

In this paper, we present an introduction to the core science themes developed by the Mauve Science Team. Sec.~\ref{sec:PlatformPayload} introduces the specifications of the Mauve satellite, Sec.~\ref{commcal} presents the commissioning and calibration plans while Sec.~\ref{sec:mauvesciprog} provides the breakdown of the science areas Mauve will cover.

\section{Platform and Payload}
\label{sec:PlatformPayload}

The satellite, a 16U smallsat, designed and built by BSSL’s industrial partners, C3S and ISISpace, houses a 13 cm Cassegrain telescope and two fibre-fed spectrometers for redundancy. Each spectrometer contains a CMOS linear array detector, covering the 200 to 700 nm range in a single exposure. Table \ref{tab:satellite} provides a breakdown of the technical specifications of the Mauve spacecraft and its payload. More details on the payload specification and assembly of the Mauve satellite can be found in Stotesbury et al. (in prep.). 

As final operational performance will only be known post-commissioning, the BSSL team and the Mauve Science Team are considering multiple performance scenarios to prepare the science programme and optimise the observation schedule. MauveSim \citep{10.1093/rasti/rzaf045}, an end-to-end simulator, was developed to assess the science capabilities of the satellite. The simulations presented here are derived from the most recent ground-based testing and expected in-orbit performance, with appropriate margins applied. The detector temperature, for instance,  directly influences the thermally induced noise on the detector and the system's overall performance.
Following commissioning, MauveSim will be revised and updated to reflect the final in-flight performance. Pending commissioning and final performance, some Mauve science themes may be updated to account for spacecraft in-orbit behaviour.  

\begin{table}
\caption{Mauve satellite technical specifications.}\label{tab:satellite}
\centering
\small
\renewcommand{\arraystretch}{1.05}
\begin{tabular}{L{2.65cm}L{5.15cm}}
\toprule
\textbf{Item} & \textbf{Specification} \\
\midrule
Telescope & 13 cm Cassegrain \\
Spectral Range & 200–700 nm \\
Spectral Resolution & 10.5 nm (\textit{R} = 20-65) \\
Sky Coverage & –46.4 to 31.8 deg (ICRS coord. ep=J2000) \\
Orbit & Sun-synchronous LEO, 510 km, LTDN 10:00 \\
Pointing Solution & High-performance star tracker and gyro \\
\bottomrule
\end{tabular}
\end{table}

\section{Commissioning and Calibration}
\label{commcal}
Mauve is equipped with a low-resolution spectrometer, resulting in each detector resolution element spanning approximately 40 pixels. This configuration leads to significant oversampling of the instrument line function -- a matrix that describes photon energy redistribution, i.e., the probability of detecting an incident photon at any given pixel. Due to this oversampling, each pixel collects photons from a range of wavelengths, making it impossible to determine the exact energy of a photon striking a specific pixel.
Any observation obtained with Mauve can be represented as the matrix product of the true stellar spectrum, the telescope’s effective area (ARF, representing total detector throughput), and the instrument line function (RMF). However, this matrix operation is non-reversible, meaning that the true stellar spectrum cannot be directly separated from the instrumental effects to recover the intrinsic stellar flux in physical units.
Because all flux calibration information is contained within the ARF and RMF, calibrating Mauve involves adjusting these arrays such that the observed spectrum of a stable reference star with a well-known spectrum matches the simulated spectrum of the same source. When the observed and simulated spectra of the reference star match with each other, the calibration is deemed successful. If discrepancies remain, the ARF and RMF are iteratively refined until a satisfactory match is achieved.
During the commissioning phase and routinely throughout nominal operations, BSSL will generate and update the calibration files required by users to perform accurate data analysis.

{The pointing stability was simulated and modelled by ISISpace, but is a primary analysis of the commissioning stage.  Prior to launch the relative pointing error (RPE) of the system was estimated to be approximately 12” across a 10 s window. This information has been incorporated into a pointing model for the system, enabling time-domain jitter analysis. Our estimates indicate that jitter-induced noise is negligible. The fibre diameter is 230 $\mathbf{\mu}$m, corresponding to a half-cone FoV of approximately 47”.}

A set of predefined standard calibration targets has been selected and will be observed periodically to monitor any variations in sensitivity and performance throughout Mauve’s operational lifetime (Saba et al. in prep.). The current list of calibration targets includes bright, well-characterised, and photometrically stable stars, primarily A- and B-type stars that emit sufficient flux within the Mauve spectral range. These targets are chosen for their stability (they are non-pulsating and non-variable), high brightness in the Mauve band, ability to achieve a high signal-to-noise ratio (S/N) within seconds, and to avoid detector saturation during short exposures. Additionally, non-variable solar-type stars have been included among the calibration targets to improve calibration in the redder wavelengths.

{To establish the initial post-launch pointing calibration, we selected Jupiter due to its exceptional brightness and large angular size. The planet enabled a first-order determination of the initial misalignment between the star tracker attitude solution and the telescope boresight. Subsequently, bright point sources (stars) were used to confirm and refine the misalignment identified from the Jupiter observations.}

A comprehensive commissioning and calibration report will be released following Mauve’s commissioning period, summarising the lessons learned during spacecraft commissioning and outlining the achievable calibration performance of the instrument. {The commissioning phase is currently expected to last up to four months. However, BSSL will review this timeline with its industry partners and announce any schedule updates as needed.}

\section{The Mauve Science Programme}
\label{sec:mauvesciprog}
\subsection{Mauve Science Planning}


While \ac{bssl} is responsible for the development and operation of the satellite platform, as well as implementation of the observation schedule, the Mauve Survey Programme is designed as a collaborative, member-driven scientific initiative. Survey members contribute to the definition of science themes, objectives, target lists, and observational strategies through a structured planning process conducted in coordination with \ac{bssl}.

{The programme is led by the Mauve Science Team, a consortium of scientists from institutions across Europe, Asia, and North America. The Science Team was established during the survey design phase, prior to Mauve launch, through a membership subscription model under which participating organisations joined the collaborative survey programme. Institutions that joined at this early stage also joined the Science Team and contributed to survey planning and prioritisation activities.}

The participating member institutions include:

\begin{itemize}
    \item Boston University 
    \item Columbia University
    \item INAF - Osservatorio Astrofisico di Arcetri 
    \item Konkoly Observatory
    \item Kyoto University
    \item Maynooth University
    \item National Astronomical Observatory of Japan 
    \item Rice University 
    \item Vanderbilt University
    \item Western University
\end{itemize}

Science planning is conducted on an annual cycle. The core science themes for Year 1 are now largely established, with observations scheduled to take place throughout 2026. However, the programme is dynamic and adaptable, and the Science Team will periodically review and update the observation schedule during the first year, incorporating new targets and refining strategies as needed. 

Although the Year 1 Science Team has now been closed, new institutions and researchers are expected to join as members. These new members will have equal access to Mauve data and the opportunity to submit observation requests for the Mauve filler programme for year 1. Years two and three of the Mauve survey are still open, and planning for year 2 is expected to commence in April 2026. 

In total, 5000 hours have already been allocated to the collaborative Mauve Science Programme for Year 1, and distributed among the current themes. In addition, \ac{bssl} will retain a reserve pool of observing hours to accommodate performance margins, future requirements and new science opportunities. 

\subsection{Science Overview}
\begin{figure}
    \centering
    
    \begin{subfigure}{0.55\textwidth}
        \centering
        \includegraphics[width=\linewidth]{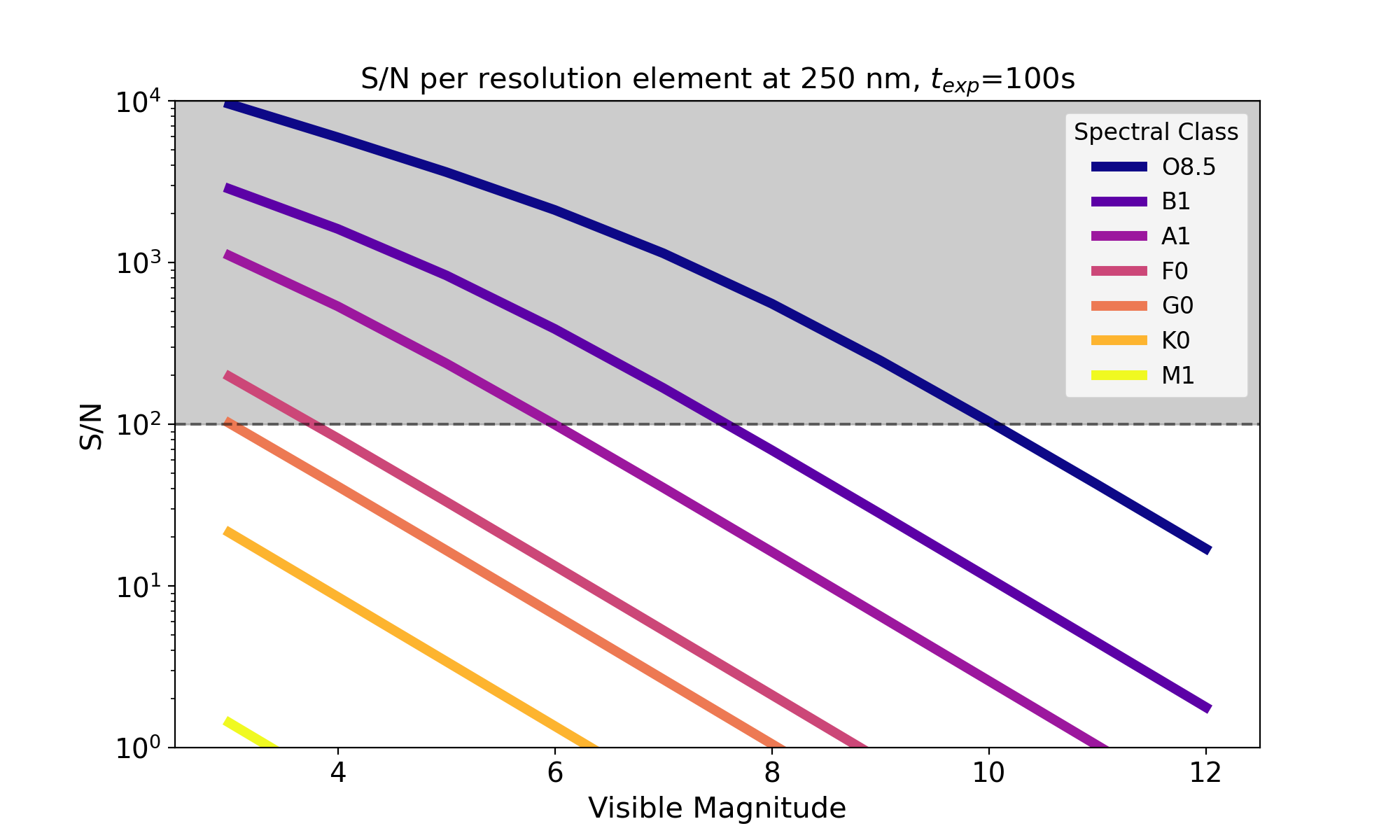}
        \label{fig:sensitivity250}
    \end{subfigure}
    \hfill
    \begin{subfigure}{0.55\textwidth}
        \centering
        \includegraphics[width=\linewidth]{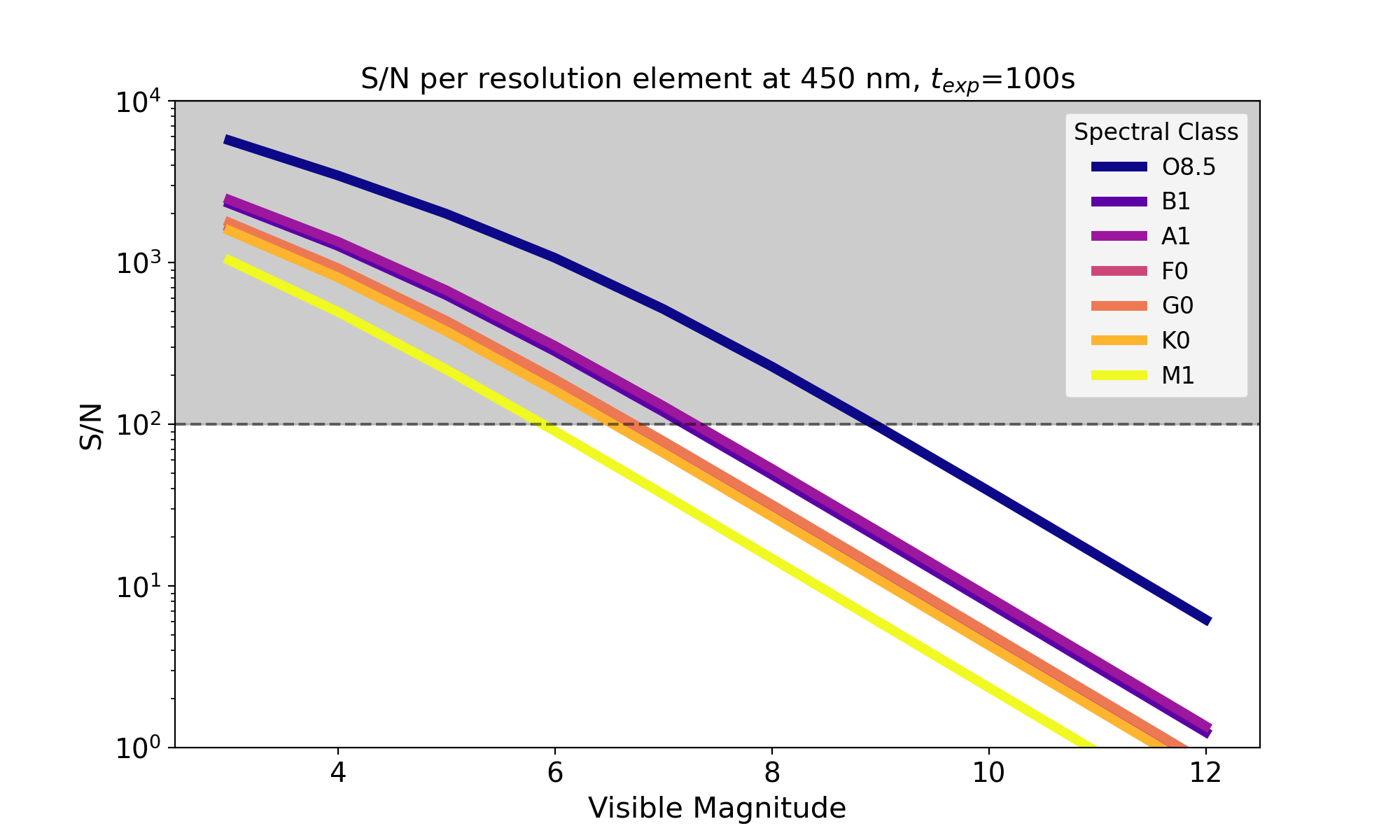}
        \label{fig:sensitivity450}
    \end{subfigure}
    \hfill
    \begin{subfigure}{0.55\textwidth}
        \centering
        \includegraphics[width=\linewidth]{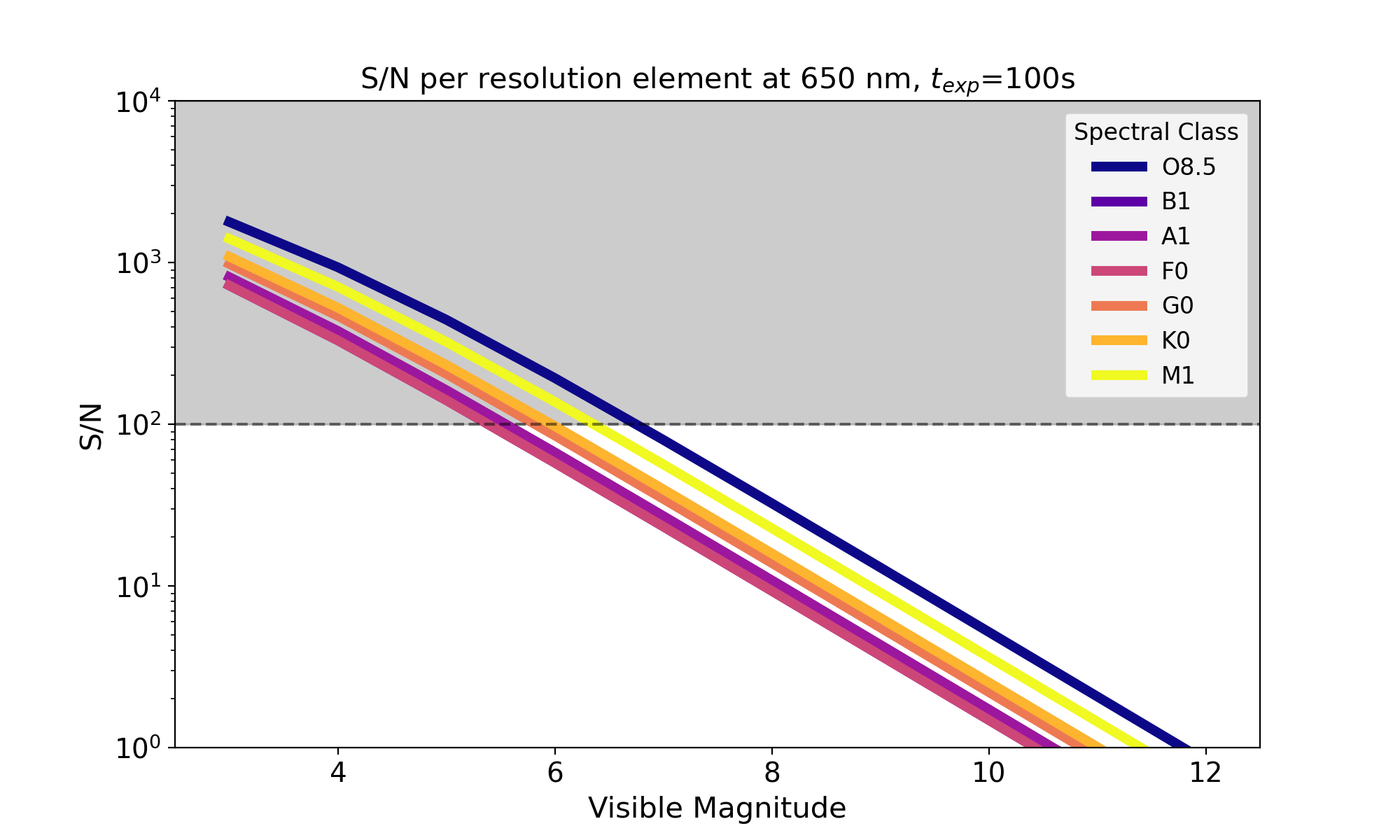}
        \label{fig:sensitivity650}
    \end{subfigure}
    
    \caption{Expected Mauve S/N per resolution element (10.5 nm) at 250 nm, 450 nm and 650 nm for a 100 s exposure time for different stellar types. The performance at the faint end is likely to be limited by the capacity of the spacecraft to track the target, which depends on factors such as stellar type and the presence of other sources in the field. Performance may also be constrained by systematics (e.g., noise saturation) at S/N per resolution element higher than 100. This is indicated in the plots as a shaded grey region above the dashed line.}
    \label{fig:sensitivityplots}
\end{figure}

{Pre-flight constraints on the field of regard (FoR) limit Mauve observations to targets within the declination range of +31.8$^{\degree}$ to -46.4$^{\degree}$; sources outside these bounds are currently inaccessible. In addition, pre-flight analyses were performed to estimate the telescope sensitivity limits and the corresponding target magnitude thresholds across different wavelength bands. The sensitivity plots in Fig.~\ref{fig:sensitivityplots} show the expected signal-to-noise ratio (S/N) per 10.5 nm bin (one resolution element) as a function of visible magnitude for several wavelength bands. These predictions are based on pre-launch performance models and ground-based characterisation of the dark current noise (450 counts s$^{-1}$) and overall payload efficiency (see \cite{mauvesim}). These sensitivity estimates are being progressively refined through ongoing commissioning activities and in-flight performance characterisation. Updated sensitivity curves, reflecting the measured on-orbit performance of the instrument, will be released following completion of the commissioning and calibration phase.}

\begin{figure*}
    \centering    \includegraphics[width=1.0\linewidth]{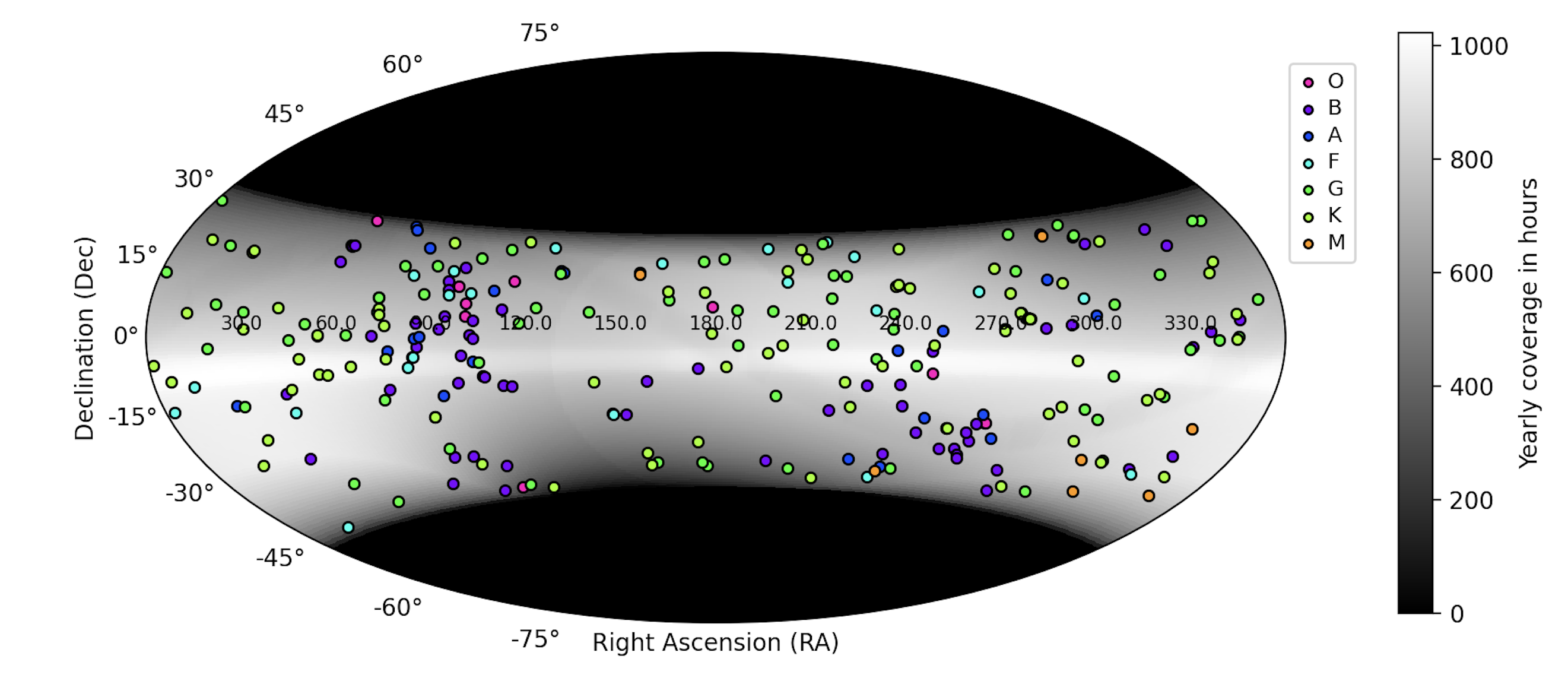}
    \caption{Candidate target list coloured by spectral type across the Mauve field of regard.}
    \label{fig:stellarMassTemp}
\end{figure*}

An initial programme of 10 core science themes has been developed by the Science Team in the first year, with approximately 290 candidate stars identified by these themes (Fig.~\ref{fig:stellarMassTemp}). Table~\ref{tab:mauve_programmes} summarises the science themes and the chosen observational strategy. 

The selected science themes can be broadly categorised into the following research areas: stellar activity and variability, host-star/planet interaction, hot stars, and exotic populations in binaries. Specific themes can then also be grouped by three main observation strategies, i.e. long-duration monitoring programmes, short-cadence single snapshots, and short-cadence monitoring with repeats. These categories determine scheduling strategy and observation cadence. Long-duration monitoring {(10-250 hours)} is typically focused on tracking stellar variability on specific targets over time. The themes using this observation strategy include studies of stellar activity and variability, specifically to understand the flare occurrence rates and to constrain the signatures of \acp{cme} through flare-associated dimming. 

The second category comprises short-cadence {(<5 hours)} single snapshots, typically associated with large-scale statistical programmes observing many targets, or small groups of targets observed repeatedly. The themes related to short-cadence single-snapshot observations include investigations of exoplanet host stars and future \ac{hwo} targets,
the study of quiescent \ac{nuv} emission in low-mass stars to constrain the evolution of rotation and magnetic activity in Sun-like (and lower-mass) stars, and the classical Be star survey, designed to explore the mechanism that causes disk ejection. 

Finally, the last observation strategy is short-cadence monitoring with repeat observations. These enable characterisation and assessment of temporal evolution. The science themes that have adopted this observational strategy include young planet-host stars, accretion variability and dipper/burster behaviour in Herbig Ae/Be stars, and studies of binary stars in exotic populations. These studies will elucidate the transition process from magnetic to non-magnetic accretion, and how key accretion and disk mechanisms evolve with increasing stellar mass in the context of star and planet formation. Additionally, Mauve will expand our understanding of binary evolution, revealing how a two-star system creates unique evolutionary pathways that are poorly understood.

\begin{table}
\caption{Mauve science themes and corresponding observational programme types.}
\centering
\small
\renewcommand{\arraystretch}{1.05}
\begin{tabular}{L{3.45cm}L{4.5cm}}
\toprule
\textbf{Science Theme} & \textbf{Observational Programme Type} \\
\midrule
M-dwarf Flares & Long-duration monitoring. \\
Superflares on Young Sun-like Stars & Long-duration monitoring. \\
Stellar CMEs & Long-duration monitoring. \\
Quiescent UV Emission in Low-Mass Stars & Short-cadence, single snapshots.\\
Future HWO Targets & Short-cadence, single snapshots. \\
Young Planet Hosts & Short-cadence monitoring with repeats.  \\
Classical Be Star Survey & Short-cadence, single snapshots. \\
Accretion Variability (Herbig Ae/Be) & Short-cadence monitoring with repeats. \\
Dipper/Burster Behaviour (Herbig Ae/Be) & Short-cadence monitoring with repeats. \\
\multirow{2}{*}{Binaries in Exotic Populations} & Short-cadence monitoring with repeats. \\
& Short-cadence, single snapshots. \\
\bottomrule
\end{tabular}
\label{tab:mauve_programmes}
\end{table}

\subsection{Understanding Stellar Variability and Activity}
\label{sec:sva}

Stellar flares and \acp{cme} are energetic manifestations of stellar magnetic activity caused by the rapid and sudden release of magnetic energy via magnetic reconnection. In the case of the Sun, flares are sometimes accompanied by an increase in solar energetic particles, exhibiting a rapid increase in electromagnetic radiation from X-ray to radio; in some cases, they are also accompanied by \acp{cme} which release large amounts of plasma into space \citep{2011LRSP....8....6S}. While the impact of solar flares and \acp{cme} on planetary space weather is well established \citep{2017LRSP...14....5K}, studies of flares and \acp{cme} for other stars and their impact on the planets orbiting around them remain underdeveloped. The goal of this science theme is to utilise Mauve's time-domain  capabilities through long-duration monitoring of stars to study the impact of stellar variability and activity on a planet's chemistry and atmospheric escape. In addition, the short-cadence, single snapshots of low-mass stars will be utilised to quantify quiescent UV emission on low-mass stars. Through these observational strategies, these science themes aim to study the nature of \ac{nuv} and optical continuum radiation from stellar flares on M-dwarfs and young G/K-dwarfs (Sun-like stars), probe stellar \acp{cme} via \ac{uv} dimming signatures, and constrain the evolution of rotation and magnetic activity in Sun-like and lower-mass stars.

\subsubsection{Radiation mechanisms of stellar flares on M-dwarfs and young Sun-like stars}

\begin{figure}
    \centering
    \includegraphics[width=1\linewidth]{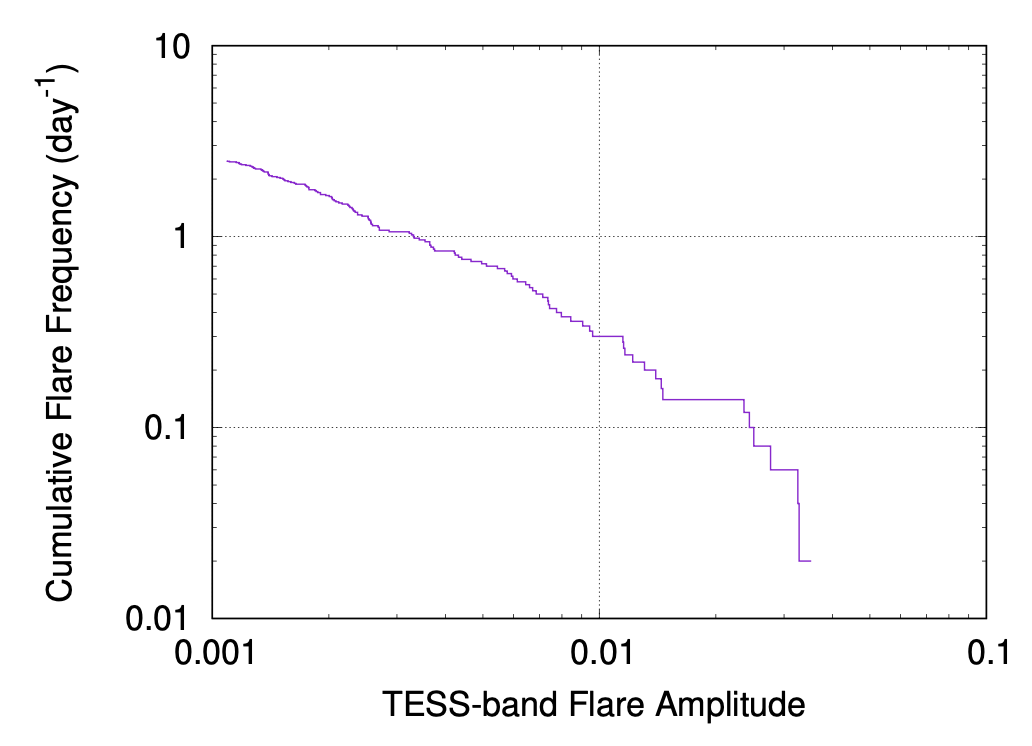}
    \caption{Cumulative flare frequency {for AU Mic} as a function of the flare amplitude in the TESS bandpass (600-1000 nm). The flare data are taken from \citet{2023ApJ...948...64I}.}
    \label{fig:maehara1}
\end{figure}

The UV and X-ray radiation from stellar flares is thought to influence planetary atmospheres and surface environments, but its actual contribution is not well justified. 
Young Sun-like (G/K-type) stars, particularly those younger than $\sim$600 Myr, exhibit extremely active magnetic behaviour, producing ``superflares" {with energies exceeding $10^{33}$ erg above their quiescent emission nearly daily}, as revealed by Kepler and TESS observations \citep[e.g.][]{2012Natur.485..478M,2022NatAs...6..241N}.
This suggests that such energetic flares on the early Sun could have significantly impacted atmospheric escape and chemistry on early Venus, Earth, and Mars when life may have first emerged. 
Likewise, exoplanets in the habitable zones of M dwarfs orbit very close to their host stars and may be subjected to repeated flare-driven UV irradiation, potentially altering their atmospheric chemistry \citep{2021NatAs...5..298C}. 
Thus, understanding the UV–optical continuum radiation of flares across stellar types is central to both exoplanetary habitability and early solar system evolution.

Despite its importance, the physical origin of the flare UV–optical continuum remains poorly understood. Many previous studies assumed an optically thick ``10,000 K blackbody" continuum when estimating flare bolometric energies, but this assumption is based on limited observations of M-dwarf flares \citep{1991ApJ...378..725H} and is unlikely universally valid. 
Solar flares frequently exhibit lower effective temperatures \citep[][]{2017ApJ...851...91N}, whereas some M-dwarf flares display much higher temperatures ($\sim$20,000–30,000 K; \citealt{2020ApJ...902..115H}). 
Moreover, both solar and stellar flares sometimes show strong Balmer jump features and optical–UV continuum shapes indicative of optically thin or multi-temperature components \citep{2013ApJS..207...15K}. Recent numerical models also suggest that optically thin radiation may be an origin of superflares on Sun-like stars \citep{2024MNRAS.528.2562S}. 
Because bolometric energy and UV flux estimates in Kepler/TESS studies depend directly on the assumed continuum model, any systematic deviation from the conventional blackbody assumption can lead to order-of-magnitude errors, profoundly affecting solar, stellar, and exoplanetary studies.

A unified understanding of the continuum radiation in both M-dwarf and young G/K-type stellar flares would therefore have wide-ranging implications. Improved UV flux estimates will refine atmospheric photochemistry and escape models for terrestrial exoplanets around M dwarfs. Re-assessing the UV output of the early Sun will advance our understanding of the atmospheric evolution of Venus, Earth, and Mars. Furthermore, identifying whether flare continuum emission arises from optically thin hydrogen recombination, optically thick Balmer/Paschen continua, or blackbody-like emission at high temperature will fundamentally revise our understanding of flare heating and radiative processes. These improvements will recalibrate the bolometric energies of the $\sim$millions of flares detected in Kepler/TESS light curves, providing a physically grounded interpretation for stellar activity across different spectral types. Within this broader context, the plan is to carry out UV monitoring observations for M-dwarf and young Sun-like stars.

Objective 1: AU Mic presents a unique opportunity to characterise the continuum properties of M-dwarf flares. Thanks to Mauve’s seamless 200–700 nm wavelength coverage, a time-resolved UV–optical spectra can be obtained. The cumulative flare frequency distribution of AU Mic derived from TESS Cycles 1 and 3 \citep[Figure~\ref{fig:maehara1};][]{2023ApJ...948...64I} suggests that a 10-day observation should detect at least one flare with a TESS-band amplitude of $\sim$2.4\% and several events with amplitudes around 1\%. Since flares with amplitudes $>$0.1\% occur at a rate of $\sim$3 day$^{-1}$ and typically last less than 30 minutes, over 90\% of time-resolved spectra will be obtained during quiescent periods. By summing more than 360 quiescent spectra (corresponding to $>$1800 sec exposure), a high S/N pre-flare spectrum will be constructed to enable reliable flare-only spectral extraction.

\begin{figure}
    \centering
    \includegraphics[width=1\linewidth]{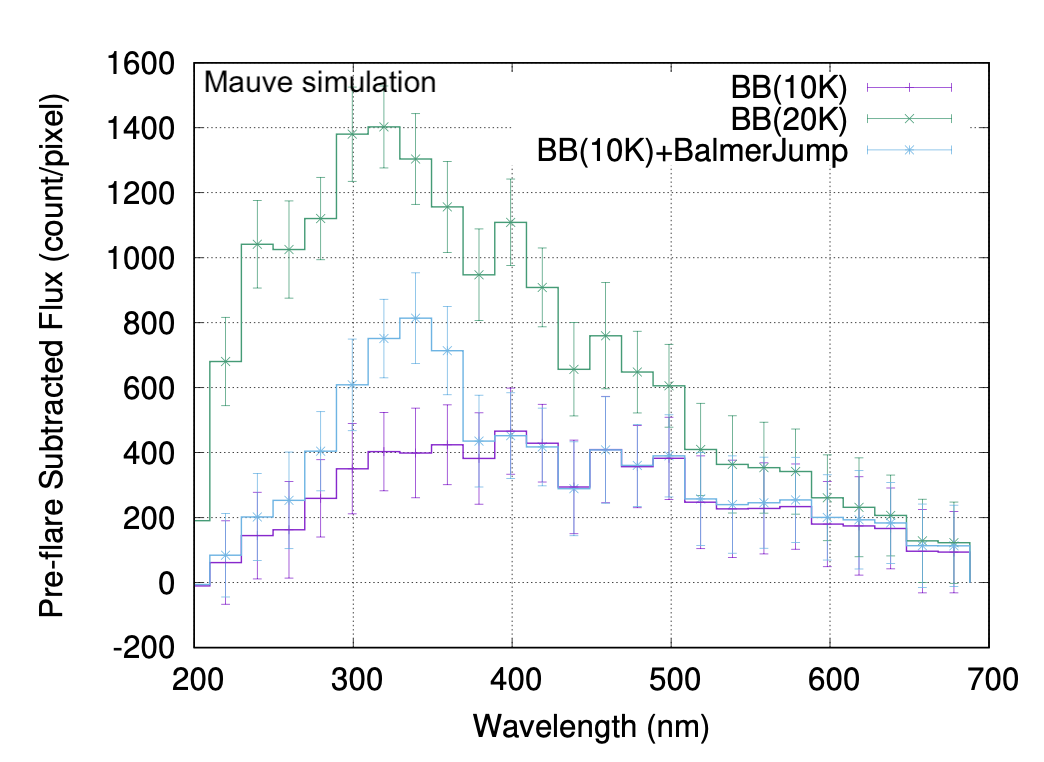}
    \caption{Comparison of simulated spectra based on different models for flare continuum. The flare amplitude for all models is 2.4\% in the TESS bandpass. The error bars represent three standard errors of mean for each bin.}
    \label{fig:maehara2}
\end{figure}

{Simulated Mauve spectra of a flare with a 300-sec exposure indicate that, after binning over 40–80 pixels (10–20 Å), the blackbody continua with T$_{\rm eff}$=10,000 and 20,000 K can be distinguished. In addition, the difference in the simulated spectra for the blackbody continua T$_{\rm eff}$=10,000 and for the continua composed of a $\sim$10,000 K blackbody plus an optically thin Balmer-jump component with a Balmer jump ratio of $\sim$3 suggests that the NUV excess flux in the wavelength shorter than the Balmer limit (346.6 nm) due to the Balmer-jump component can be detectable with Mauve.} See Figure \ref{fig:maehara2} for the model calculation. The minimum success criterion of this theme is to detect at least one flare and measure its effective temperature and Balmer jump ratio, while full success consists of detecting $>$3–4 flares and investigating the diversity of their NUV continua. If a superflare occurs by chance, Mauve’s short exposure capability (30–60 sec) will allow us to resolve temporal changes in the continuum throughout the flare, providing unique constraints on the heating processes governing powerful M-dwarf flares.

Objective 2: Young Sun-like (G/K-type) stars such as V889 Her and $\epsilon$ Eridani offer an equally important window into the continuum properties of stellar superflares. These stars are rapid rotators with strong magnetic fields and produce frequent superflares with energies $\gtrsim 10^{34}$ erg. {Although the associated brightening in the optical band is only $\sim$0.5–1\%, this does not reflect the total energetic output of the event. While cool stars have photospheric temperatures of roughly 3000-6000 K, the flare emission can be characterised by an effective temperature of $\sim$10,000 K (or higher). This naturally leads to a stronger relative enhancement at shorter wavelengths (UV and X-rays), while the optical contributes only a small fraction to the bolometric flare energy. The low optical contrast makes ground-based spectroscopic observations nearly impossible.} TESS observations indicate that V889 Her flares of this scale occur every 1–2 days and last tens of minutes \citep{2025ApJ...993...80N}. Simulations assuming {a 1\% increase in flux relative to the quiescent emission} ($\approx2\times10^{34}$ erg) and a 5-minute Mauve exposure show that spectra binned to 25–50 nm can distinguish optically thin continua—characterised by a pronounced Balmer jump—from 10,000–20,000 K blackbody continua. 
Thus, observations are planned for V889 Her for 100--380 hrs considering the high flare rate in TESS.
$\epsilon$ Eridani shows only a modest flare rate in TESS photometry, yet HST FUV observations reveal that it produces flares far more frequently \citep{2022ApJ...936..170L}. Although Mauve's sensitivity is more limited than that of HST, the largest flares detectable by Mauve are expected to occur roughly once or twice per day. Therefore, the plan is to conduct 50–100 hours of observations of $\epsilon$ Eridani to capture these events.

As a goal, by combining Mauve observations with analytical continuum diagnostics \citep{2024MNRAS.532L..56H} and state-of-the-art 1D hydrodynamic models such as RADYN \citep{2020PASJ...72...68N,2024ApJ...969..121K}, the plasma temperature, column density, and the properties of nonthermal electron/proton beams responsible for heating will be inferred. These results will enable the construction of an observation-based spectral model for (super-)flares on M-dwarfs and Sun-like stars.

\subsubsection{Probing coronal mass ejections through UV dimming signatures}

\begin{figure}
\includegraphics[width=1\linewidth]{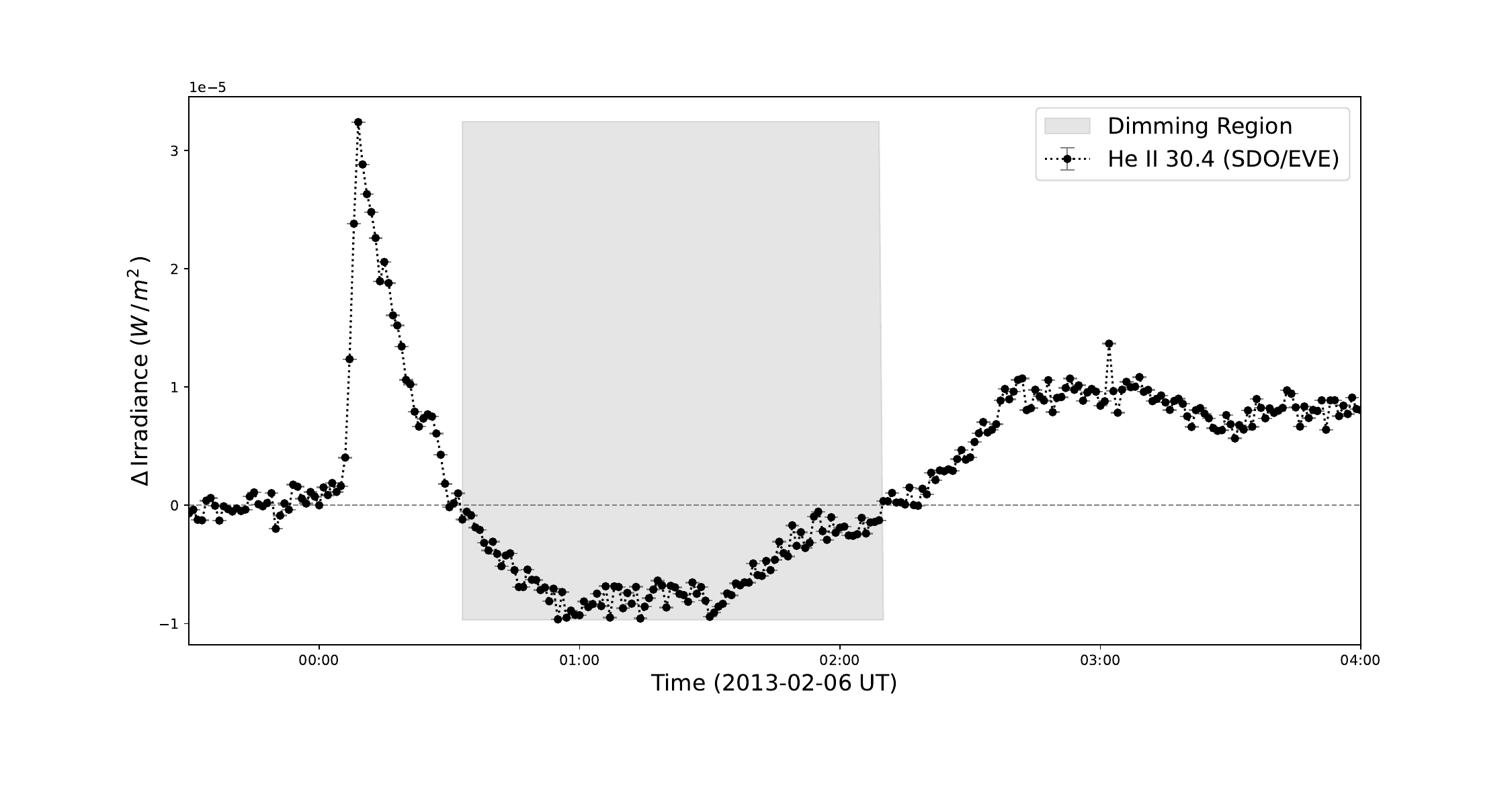}
\caption{Post-flare coronal dimming associated with a CME detected in the Sun{-as-a-star} He II 30.4 nm line from SDO/EVE. The shaded region indicates the interval of CME-related dimming. While the He II line does not lie in the wavelength region covered by Mauve, the example illustrates that post-flare coronal dimming is associated with CME-related mass depletion. We expect analogous stellar coronal dimming events within the wavelength coverage of Mauve.
\label{fig:dimming}}
\end{figure}

In contrast to stellar flares, confirmed stellar \acp{cme} beyond the Sun remain exceedingly rare due to the lack of spatially resolved diagnostics. A promising indirect method is the detection of flare-associated dimming signatures, analogous to solar coronal dimming. These dimmings may arise either from \emph{emission dimming}, where plasma is evacuated from the stellar atmosphere, or from \emph{obscuration dimming}, where cool filament or prominence material blocks background radiation \citep[e.g.,][]{2014ApJ...789...61M, 2025ApJ...987L..22C}. 

A particularly valuable solar diagnostic is the He~II 30.4~nm line, formed in the transition region at $\sim$$5\times10^{4}$ K. Sun-as-a-star analyses with SDO/EVE and GOES/EUVS have shown that this line exhibits measurable obscuration dimmings during filament eruptions leading to \acp{cme} \citep{2024ApJ...970...60X}. {Figure~\ref{fig:dimming} shows that, following an eruptive flare, the Sun-as-a-star (full-disk integrated) light curve of the He II 30.4 nm line exhibits a CME-associated dimming.} 
This dimming corresponds to the transit of erupting filament material across the solar disk, which significantly attenuates the background emission. The amplitude and duration of this event demonstrate the sensitivity of transition-region diagnostics to CME-related obscuration and provide a benchmark for what could be detectable in stellar light curves. Beyond the transition region, dimming has also been observed in chromospheric diagnostics: for instance, \citet{2007SoPh..240...77J} reported H$\alpha$ dimmings associated with filament eruptions, further supporting the notion that multi-wavelength dimming phenomena are robust CME tracers across atmospheric layers. Together, these results reinforce the physical link between post-flare dimming signatures and CME mass budgets.

Building on this solar benchmark, Mauve’s high-cadence spectrophotometry offers a unique opportunity to extend dimming-based CME detection to other stars. While the He II 30.4 nm line cannot be detected by Mauve, it is a suitable proxy for dimming tracers within stellar chromospheres. Several spectral diagnostics within Mauve’s bandpass probe chromospheric and transition-region layers at $10^{4}$--$10^{5}$~K, making them conceptually analogous to He~II 30.4~nm {(e.g. \citep{2025LRSP...22....2V,2022NatAs...6..241N,1990MNRAS.244..291D})}. These include the Mg~II h/k resonance lines (2796/2803~\AA), the Ca~II H\&K lines (3934/3968~\AA), and the Balmer series (e.g., H$\alpha$ 6563~\AA, H$\beta$ 4861~\AA). Each is sensitive to plasma density variations and obscuration by cool erupting material, with line cores expected to show the strongest dimming signatures. By monitoring these diagnostics after energetic flares, Mauve can search for persistent flux decreases beyond the flare's impulsive phase. The wavelength dependence and timing will make such signatures compelling indirect evidence of stellar \acp{cme}.

Long-duration spectroscopic monitoring of active stars with high flare rates will be observed to ensure a statistically significant number of post-flare intervals during extended observing campaigns. By constructing Sun-as-a-star style light curves from Mauve spectra and applying solar-inspired analysis techniques, 
a systematic search for post-flare dimmings in Mg~II, Ca~II, and Balmer lines will be carried out. Detecting such signatures would provide strong evidence for \acp{cme} on active stars. 



\subsubsection{Quiescent UV Emission in Low-Mass Stars}

Observations of Sun-like stars throughout the 1960s showed that rotation and magnetic activity decrease over time at the same rate \citep{1972ApJ...171..565S}, with the stars’ spin-down thought to be due to angular momentum lost through magnetised winds. This coupling of rotation and the stellar magnetic field implies that observable indicators of the field’s presence, such as optical emission lines or X-ray luminosity, also weaken over time.

However, the dependence of rotation or magnetic activity on age involves poorly understood physics, such as the magnetic field geometry and the degree of core–envelope coupling (e.g., \citealt{Réville_2015}). Furthermore, theorists disagree on whether different magnetic activity indicators, which trace the heating of different atmospheric layers, should have a similar dependence on rotation. In short, many details remain unknown about how magnetic energy is distributed across stellar atmospheric layers and how that distribution changes as main-sequence stars age and spin down \citep{2019ApJ...872...17R, 10.1093/mnras/stw1936, 2013MNRAS.431.2063S}.

To constrain the evolution of rotation and magnetic activity in Sun-like (and lower-mass) stars, a sample of stars in open clusters with well-determined ages will be used \citep{Agüeros_2011, 2018ApJ...862...33A, 2019ApJ...879...49C, 2019AJ....158...77C, 2020ApJ...904..140C, 2014ApJ...795..161D, 2016ApJ...830...44N, 2017ApJ...834..176N, 2015ApJ...809..161N, Núñez_2022, 2024ApJ...962...12N}. {We will use Mauve to capture short-duration (i.e., minutes or hours) variability, typically associated with flares. Such flares will then be removed from our UV emission measurement to capture the quiescent level of emission.}

This science theme will focus on the characterisation of the transition region between the chromosphere and the corona as this hot, tenuous layer is where magnetically driven stellar UV emission originates, but how that emission varies as a function of mass, rotation, and age is poorly constrained observationally. The use of GALEX photometry data to study low-mass stars (e.g, \citep{2011AJ....142...23F, 2019ApJ...872...17R, Schneider_2018, Shkolnik_2014, 2013MNRAS.431.2063S}), will be combined with Mauve’s low-resolution UV spectrophotometry to greatly improve the ability to quantify quiescent UV emission, connect it with stellar rotation, and compare it to other indicators of magnetic activity. The primary objectives 
are: 

\begin{enumerate}
    \item to compare UV fluxes among stars exhibiting different levels of magnetic activity, thereby identifying how UV output varies with stellar properties;
    \item to evaluate the relationship between UV emission and chromospheric and coronal diagnostics, providing new insights into the influence of magnetic heating on stellar atmospheres;
    \item to validate and extend trends suggested by GALEX photometry, while assessing the added value of Mauve’s spectrophotometric coverage; and
    \item to provide input for planetary atmosphere models to study their evolution and habitability properties.
\end{enumerate}

The integrated UV fluxes over well-defined spectral bands will be extracted from the science-ready Mauve spectrophotometry. These measurements will be combined with 
existing catalogues of rotation periods, H$\alpha$ equivalent widths, and X-ray luminosities to evaluate whether Mauve UV fluxes track established magnetic activity trends, and to identify any systematic departures from those trends. By comparing UV emission with chromospheric and coronal diagnostics, the aim is to understand how magnetic heating is partitioned across stellar atmospheric layers---and how UV flux can serve as a sensitive proxy for magnetic activity, especially in cases where optical or X-ray diagnostics are unavailable or ambiguous.

Mauve data of proposed targets will allow the exploration into how UV emission varies across late F and G stars with different activity levels, rotation rates, and ages.
Furthermore, several of these targets also appear in the \ac{hwo} \ac{hpic} {\citep{2024arXiv240212414M}}. Although the focus is on the connection between rotation, activity, and age, these targets underscore the broader value of this work. Improving the understanding of the high-energy environments of solar-type stars is essential for interpreting future \ac{hwo} observations of planetary systems and their atmospheres.

\subsection{Exoplanet Hosts}

As explored in section \ref{sec:sva}, UV/X-ray radiation from stellar flares could have a significant impact on the atmospheric escape and chemistry of the orbiting planets. In the case that these flares are accompanied by stellar \acp{cme}, this can result in atmospheric erosion which further exposes the planets to stellar radiation \citep{2016ApJ...826..195K}. Thus, understanding stellar magnetic activity and variability is crucial for assessing habitability in exoplanets \citep{2016NatGe...9..452A}.

In addition to long-duration studies providing insights into M-, G-, and K-type stars, Mauve has two science themes directly related to the study of exoplanet hosts that will be undertaken within Mauve's first year: the characterisation of young planet hosts and early observations of \ac{hwo} targets.


\subsubsection{Young Exoplanet Hosts and \ac{hwo}}

\begin{figure}
   \centering
  \includegraphics[width=1\linewidth]{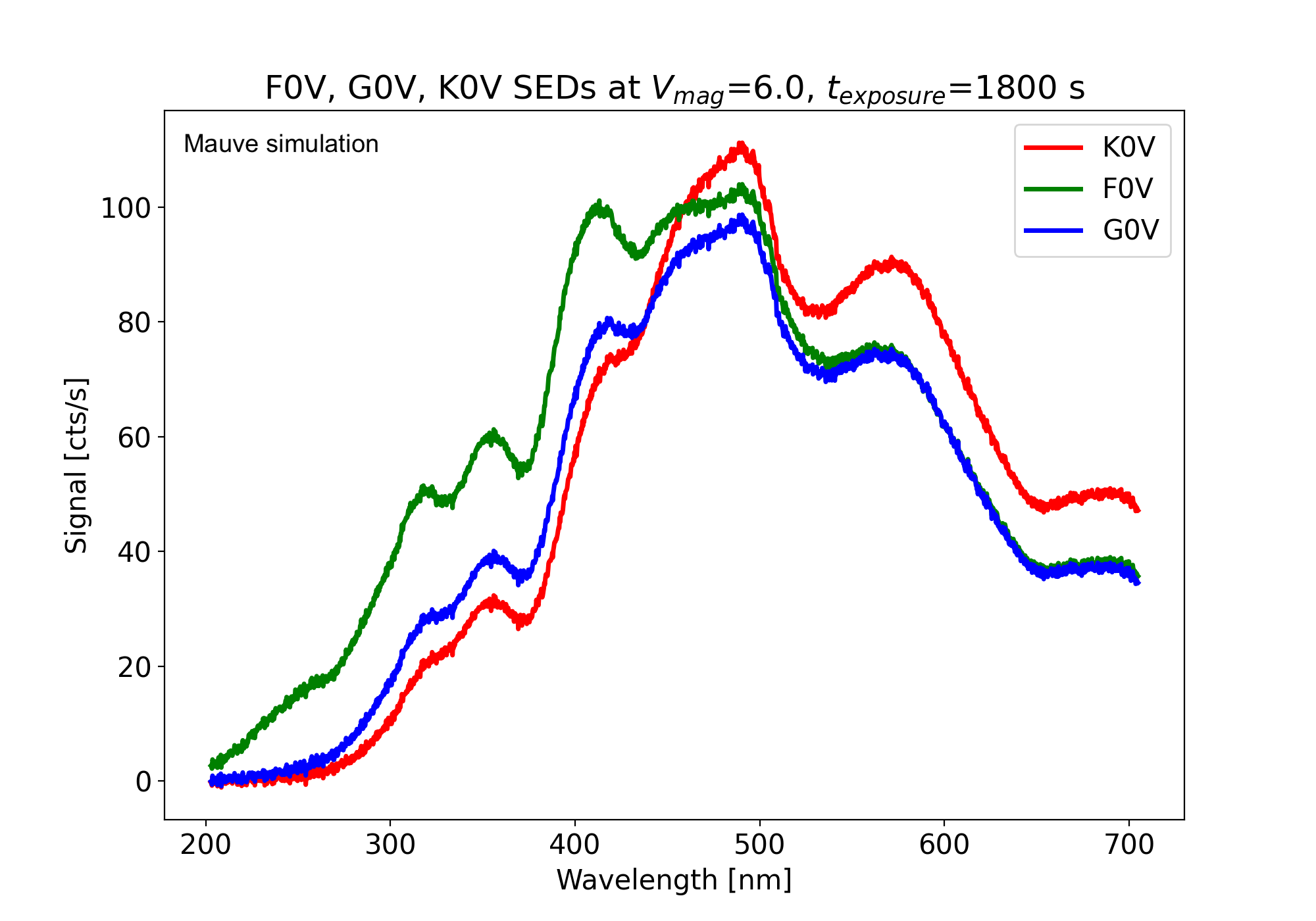}
   \centering
  \includegraphics[width=1\linewidth]{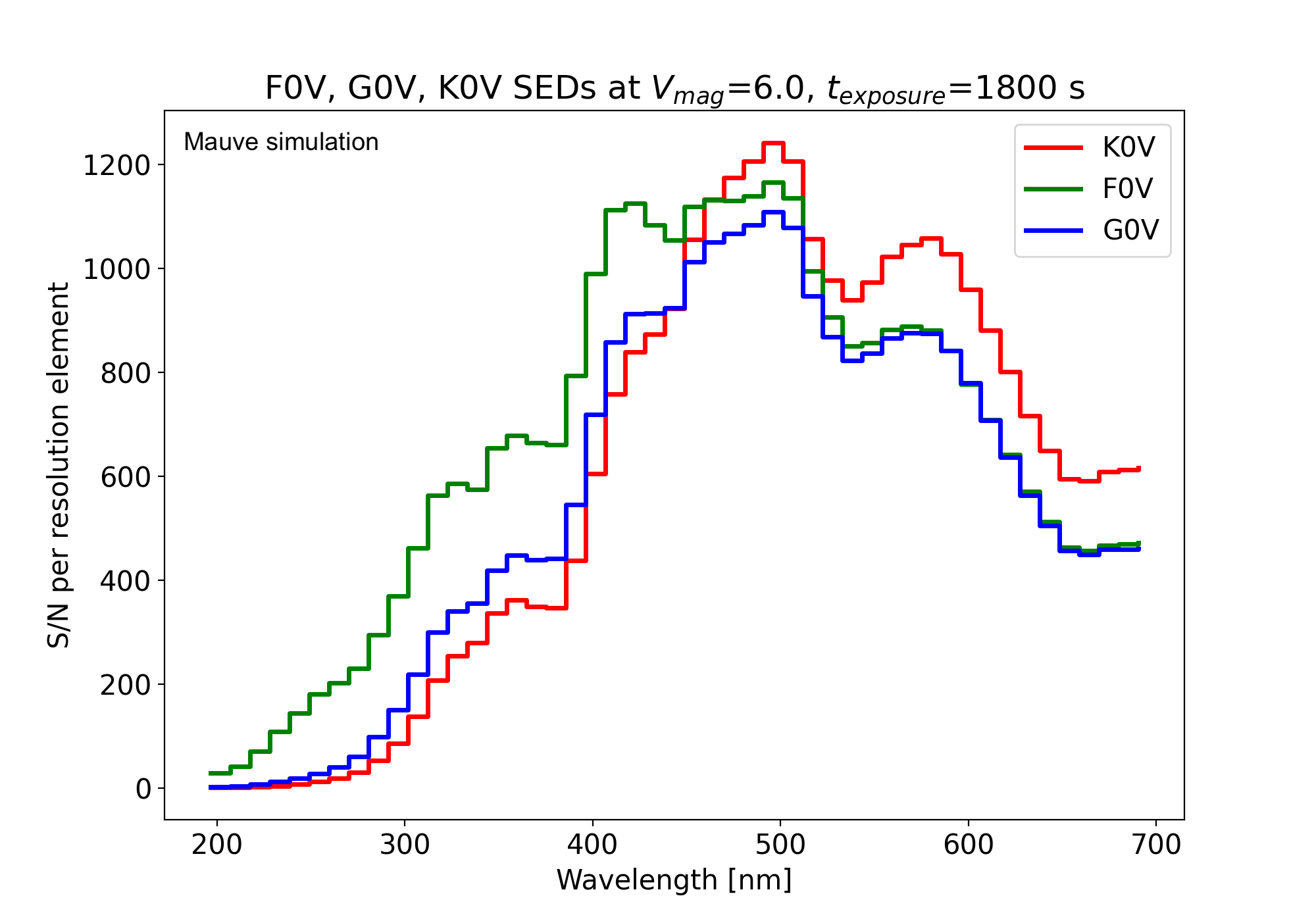}
   \caption{Simulated Mauve SEDs (upper panel) and S/N (lower panel) for example K0V, G0V and F0V stars at $V_{\mathrm{mag}}=6$ using an exposure time of 30 minutes.}
   \label{fig:girish}
\end{figure}

Observations of young planet host stars serve two purposes: complementing transmission spectroscopy observations of the young planets by providing empirical spectra, and understanding how the evolution of stellar magnetic activity manifests in the \ac{nuv}. The MUSCLES programme and its successors \citep{2016ApJ...820...89F, 2021hst..prop16701Y, 2025ApJ...978...85W} have produced panchromatic spectra of exoplanet hosts using \ac{hst} for the \ac{uv} wavelength regime and PHOENIX \citep{1993JQSRT..50..301H} models for the bulk optical and \ac{ir} emission. Stitching these two together has historically been a problem due to only some of the \ac{hst} observations including data from 250-300 nm. Mauve can fill this gap with its spectral range and replace much of the optical portion of the \ac{sed}. Figure~\ref{fig:girish} shows the expected Mauve SEDs and S/N per resolution element for example K0V, G0V and F0V stars.

Besides enabling experiments for exoplanet theorists and providing constraints for exoplanet observers, the 200-300 nm range contains many transitions formed in the chromosphere, most notably Mg II lines. Stellar activity is often parameterised using a pair of power laws as a function of Rossby number with a transition point \citep{2024AJ....168...60F}, but where this transition point is in stellar parameter space seems to be wavelength dependent (and not simply a product of different assumptions for calculating the Rossby number): coronal X-ray
breakpoints seem earlier than \ac{fuv} transition region lines or chromospheric H$\alpha$ \citep{2021ApJ...911..111P}. Mg II forms across the upper chromosphere and extends at least partially to the transition region, so studying stellar activity in the \ac{nuv} might pin down this wavelength dependence. The current candidate planet hosts span ages across 1-500 Myr, an interval that contains the breakpoints for most activity broken power law relations.

The goal of this science theme would be to measure the \ac{uv} spectra and variability of known young planet hosts, on short timescales to verify whether the measurement is being affected by a flare, and on longer timescales to account for potential activity cycle variation.


As part of the characterisation of \ac{hwo} targets theme, a sample of $\sim$40 targets has been identified within Mauve's FoR. These candidate targets have been selected from the priority A target list for the primary mission of the HWO, where they aim to directly image an Earth-sized planet in the liquid water zone of a star. 

By providing visible to NUV spectrophotometry of these targets, Mauve can fill in gaps in the \acp{sed} of these stars to inform \ac{hwo} on mission design requirements, yield estimates, and contribute to a library of panchromatic spectrophotometry for main-sequence stars. The minimum requirement is to obtain a single snapshot of the \ac{nuv} spectrum that can be published as a resource for any scientific investigation that requires the \ac{nuv} spectrum of a main sequence star as an input, but the primary intended purpose is to provide the \ac{hwo} Target Stars and Systems Working Group to inform the survey plan. 
One plausible scientific use for this library is to compare the obtained spectra to the various stellar atmosphere models in circulation, such as the PHOENIX models, to determine how they differ from ground truth data for different spectral types. Another is to inform planet atmosphere modelling codes that need \ac{nuv} spectra as inputs for work on haze production{, chemical disequilibrium,} and biosignatures.


\subsection{Classical Be and Herbig Ae/Be stars}

\subsubsection{The Classical Be star survey}
\label{sec:CBe}
\begin{figure}
    \centering
    \includegraphics[width=1\linewidth]{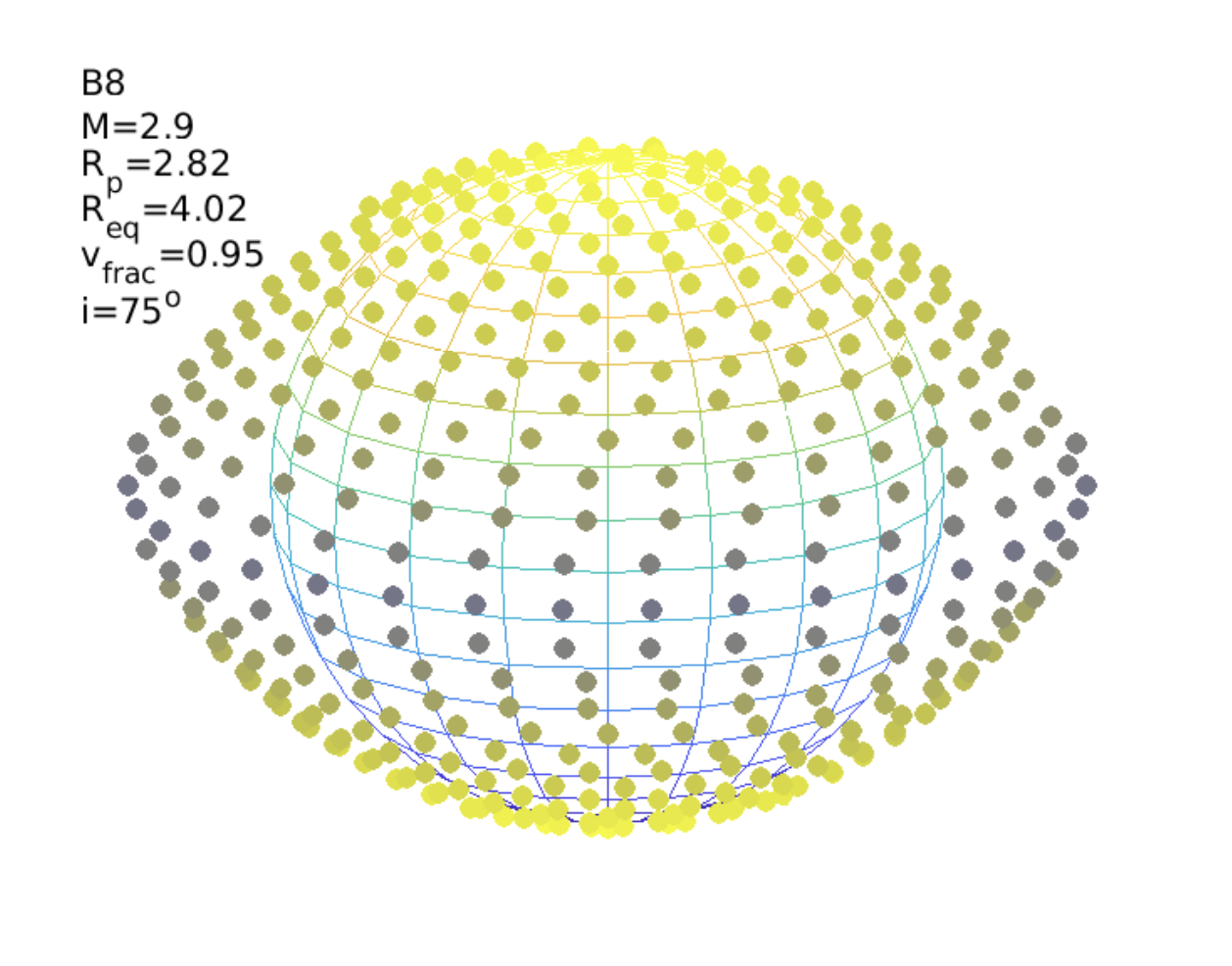}
    \caption{The distorted surface of a B8V star (mass 2.9 $M_\odot$, polar radius 2.82 $R_\odot$) rotating at 95\% of its critical velocity (363 km/s) seen at an inclination angle $i = 75^{\circ}$ ($i = 0^{\circ}$ corresponds to a star viewed pole-on). The coloured circles represent surface "patches" that can be seen by a distant observer. Gravitational darkening is illustrated by the colour of each patch, ranging from bright yellow for the hot pole ($T_{\rm eff}$ = 13,000 K) to dark grey for the cooler equator ($T_{\rm eff}$ = 8000 K). To compute the \ac{sed} seen by a distant observer, the intensity for each surface patch is computed using local values of ($T_{\rm eff}$, log $g$), and then intensities for all patches are summed, weighted by their solid angles.}
    \label{fig:sigut1}
\end{figure}
\begin{figure}
    \centering
    \includegraphics[width=1\linewidth]{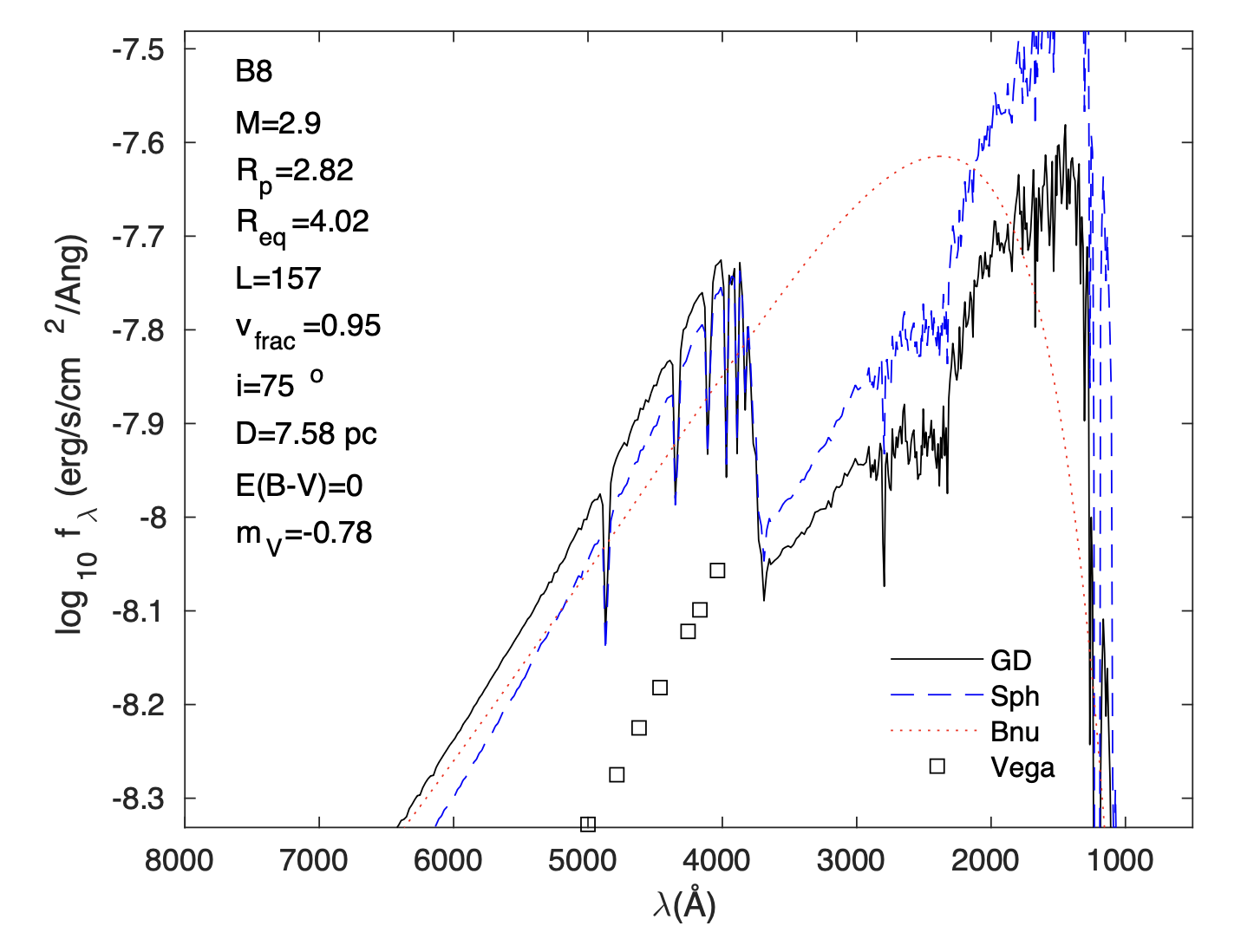}
        \caption{Simulated fluxes at the Earth for the B8V CBe star of Figure~\ref{fig:sigut1} seen at a distance of 15 pc compute with the \texttt{Bedisk} \citep{2007ApJ...668..481S} and \texttt{Beray} \citep{2018ASPC..515..213S} codes. The solid black line is the predicted optical and \ac{nuv} fluxes for gravitationally-darkened, rotationally-distorted star rotating at 95\% of its critical velocity seen at an inclination angle of $i=75^{\circ}$. The dashed blue line gives the fluxes expected for a spherical, non-rotating star. The red dotted line is the prediction of a black body at the nominal (non-rotating) $T_{\rm eff}$, and the square symbols are the calibrated Vega fluxes from \citet{gray2021observation} included for reference.}
    \label{fig:sigut2}
\end{figure}    
\begin{figure}
    \centering
    \includegraphics[width=1\linewidth]{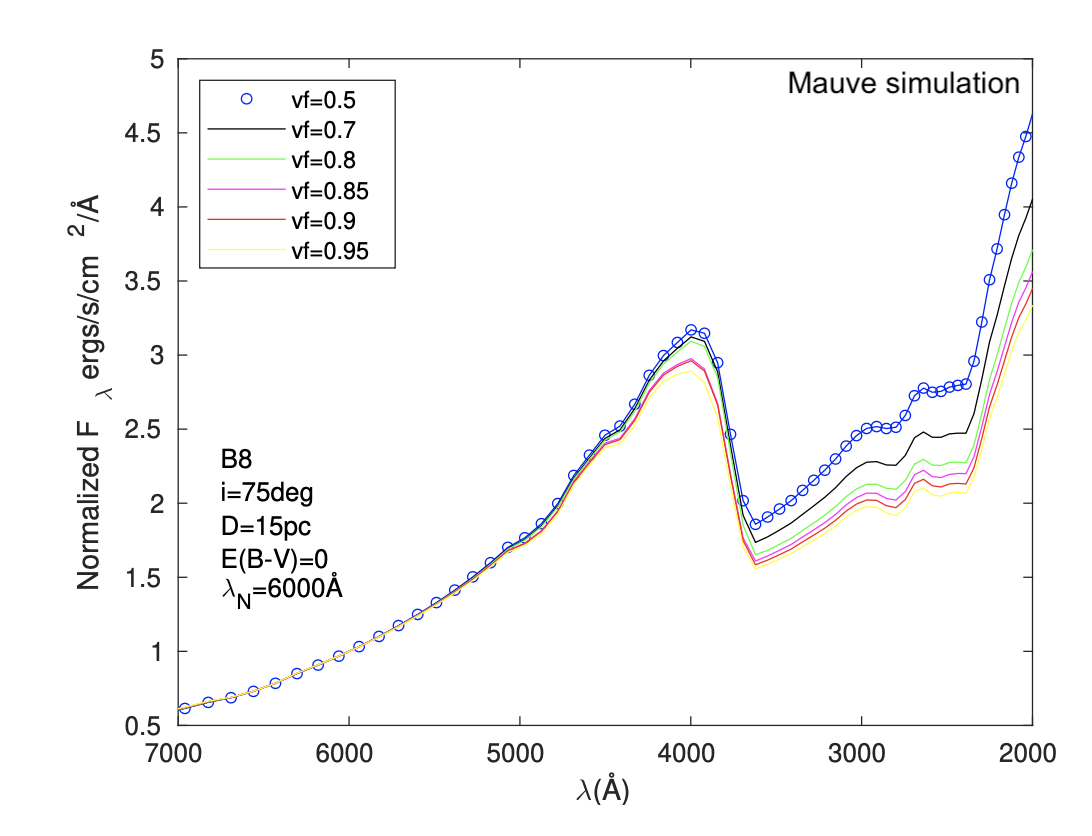}
    \caption{Predicted SEDs for the star of Figure~\ref{fig:sigut1} as a function of the stellar rotation rate $v_{\rm frac}$ (see legend). The SEDs have been convolved to the same resolution as Mauve (R= 50) and are normalised to unity at $\lambda=6000$ \AA.}
    \label{fig:sigut3}
\end{figure}

Classical Be (CBe) stars are upper main sequence, B-type stars ($2.5-20\,M_\odot$) that exhibit emission lines in their spectrum (most prominently H$\alpha$ at $\lambda = 6562.8\,$\AA), an \ac{ir} excess beyond $1\,\mu$m, and continuum polarisation at the level of approximately 1\%  \citep{2003PASP..115.1153P}. The interpretation is that \ac{cbe} stars are main sequence B stars surrounded by a thin, gaseous, equatorial disk in which the emission lines and \ac{ir} excess are formed by recombination in the disk gas \citep{2013A&ARv..21...69R}. \ac{cbe} stars are very common: it is estimated that perhaps one-fifth of all main sequence B-type stars have been \ac{cbe} stars at one time \citep{1997A&A...318..443Z}. The \ac{cbe} stars are distinct from the more familiar Herbig Ae/Be stars in that \ac{cbe} \ac{ir} \acp{sed} show no evidence for dust. While the Herbig Ae/Be stars are thought to be pre-main sequence objects still embedded in the remnants of their primordial accretion disks \citep{waters1998}, the disks of the \ac{cbe} stars are thought to be {\it out-flowing, decretion disks\/} formed by mass ejected from the central stars \citep{2013A&ARv..21...69R}. 

The central mystery of the \ac{cbe} stars is that the mechanism that causes the disk ejection is unknown despite decades of observations and modelling. Over the years, disk formation has been hypothesised to include one or more of the following: rapid stellar rotation, \ac{nrp}, magnetic fields, stellar winds, and binarity. Among this list, rapid stellar rotation has long been thought to play a key role, an idea dating back to \citet{1931ApJ....73...94S}. In this view, the \ac{cbe} stars are postulated to rotate close to their critical (equatorial) rotation speed, defined as 
\begin{equation}
    V_{\rm crit}=\left( \frac{2GM_*}{3R_{\rm{p}}} \right)^{1/2}
\end{equation}
where $M_*$ is the stellar mass and $R_{\rm{p}}$ is the polar radius\footnote{At critical rotation, the equatorial radius is $R_e=(3/2)R_{\rm{p}}$ in the Roche approximation.}. For a star rotating at its critical velocity, the effective gravitational acceleration at the equator vanishes and the material there effectively orbits the star. If a \ac{cbe} star actually rotates with an equatorial speed of $V_e$, the parameter $ v_{\rm frac}\equiv V_e/V_{\rm crit} \le 1$ controls how effective rapid stellar rotation is in aiding disk ejection. Some postulated disk ejection mechanisms depend very strongly on the value of $v_{\rm frac}$. For example, a promising candidate is \ac{nrp} \citep{2003A&A...411..229R}, yet to be effective, \ac{nrp} requires $v_{\rm frac}$ close to unity as it naturally produces flows with velocities on the order of the sound speed in the gas ($\approx\!10\,\rm km\,s^{-1}$); for these perturbations to result in the ejection of material, the star must be very near critical rotation.

Spectroscopic estimates of $v_{\rm frac}$ are complicated by the need for accurate stellar parameters and good measurements of $v\sin i$, the projected equatorial rotation velocity which can be deduced from the rotational broadening of spectral lines. For many years, measurements suggested $v_{\rm frac}\sim 0.8$, indicating that while the \ac{cbe} stars are rapid rotators, they are not {\it critical rotators}. This conclusion was revised by \citet{2004MNRAS.350..189T} who noted that gravitational darkening of the central B star due to rapid rotation can cause equatorial rotation speeds derived from $v\sin i$ measurements to be significantly underestimated. It has long been known that rapid stellar rotation leads to a distorted stellar surface and a latitude-dependent $T_{\rm eff}$ and $\log(g)$ \citep{1924MNRAS..84..665V,2011A&A...533A..43E}. For example, Figure~\ref{fig:sigut1} shows a B8V star rotating at 95\% of its critical rotational speed. The latitude-dependent $T_{\rm eff}$ is particularly important as for $v_{\rm frac}\approx\!1$, the local $T_{\rm eff}$ at the equator becomes significantly cooler, and thus the most rapidly-rotating portion of the stellar surface has its contribution to the overall spectrum diminished, with $v\sin i$ correspondingly underestimated. Despite initial enthusiasm for this idea \citep{2013A&ARv..21...69R}, subsequent detailed modelling of the optical spectra of \ac{cbe} stars including gravitational darkening has not established that the \ac{cbe} stars as a group are critically-rotating. Based on a sample of 233 \ac{cbe} stars, \citet{2016A&A...595A.132Z} found $<\!v_{\rm frac}\!>\approx 0.8$ with the distribution characterised by the very wide range $0.35\le v_{\rm frac}\le 0.95$, implying that much more energetic mechanisms are often required for disk ejection. Such mechanisms could potentially involve magnetic fields \citep{2021ApJ...921....5B}\footnote{There are currently no detections of magnetic fields in \ac{cbe} stars, despite large survey searches.} or binarity \citep{2021A&A...653A.144H}.

Despite this picture, there still is hope for the critical rotation hypothesis. One significant omission in all of the analyses above is the neglect of the circumstellar disk in the modelling of the \ac{cbe} \ac{sed} and in the extraction of the $v\sin i$ estimates. In addition, all of this work is based upon ground-based, optical spectra. Access to space-based \acp{sed} for the \ac{cbe} stars opens the possibility of determining $v_{\rm frac}$ directly from the influence of gravitational darkening on the central star's \ac{nuv} \ac{sed}.

The goal of this science theme is to determine if the \ac{uv} portion of late-type \ac{cbe} stellar \acp{sed} is consistent with the $v_{\rm frac}$ values obtained from optical spectra, or if the $v_{\rm frac}$ values need to be revised. This is a sensitive test of the models because the \ac{uv} portion of the spectrum is much more sensitive to the temperature variations across the stellar surface caused by gravitational darkening. For late-type \ac{cbe} stars, spectral types B4 and later, the \ac{nuv} fluxes accessible by Mauve are shortward of the Rayleigh-Jean’s blackbody tail and hence are very sensitive to variations in the local $T_{\rm eff}$ across the stellar surface. This is illustrated in Figure~\ref{fig:sigut2} where the flux at the Earth is simulated for the rapidly-rotating \ac{cbe} star of Figure~\ref{fig:sigut1}.
An example of the expected Mauve signal seen for the same star in Figure~\ref{fig:sigut1} is shown in Figure~\ref{fig:sigut3}. Here, the optical and \ac{nuv} \acp{sed}, convolved to a resolution of $\lambda/\Delta\lambda$ = 50, are shown for various rotational speeds ranging from $v_{\rm frac}$ = 0.50 to 0.95. Shown are relative fluxes, all normalised to unity at $\lambda$ = 6000 \AA. Interstellar reddening, E(B-V), was assumed to be zero, although the modelling can account for interstellar reddening following \citet{1999PASP..111...63F}. 

This survey aims to use Mauve to observe a sample of about 65 \ac{cbe} stars of spectral type B4 or later. These targets were selected from the list of \ac{cbe} stars from \citet{2016A&A...595A.132Z}, all of which have estimates of $v_{\rm frac}$ based on optical spectra. The filtering was done for Mauve sky visibility and the B4 spectral type limit. Each target has estimated stellar parameters ($M, R_{\rm p}, L$) and $v_{\rm frac}$ from \citet{2016A&A...595A.132Z} which can be coupled with the system viewing inclination from \citet{2023ApJ...948...34S}, and the object’s known distance and reddening, to produce a simulated \ac{sed} in absolute flux units, similar to the example given in Figure~\ref{fig:sigut2}.

In addition to the selected \ac{cbe} stars, five \ac{cae} stars will be observed: \ac{cae} stars are thought to be the cool extension of the \ac{cbe} phenomena to the cooler, A-type spectral class \citep[see][and references therein]{2025ApJ...988..129A}. These objects are rarer, and hence fainter, than the \ac{cbe} stars, yet their out-flowing disks are also thought to be driven by rapid stellar rotation \citep{2021MNRAS.501.5927A}, and their \acp{sed} may reveal this in the same way as described above for the \ac{cbe} stars.

\subsubsection{Variability of Herbig Ae/Be stars and implications for accretion and planet formation}
Herbig Ae/Be stars are an important class of young stellar objects (YSOs), as they bridge the gap between solar-type stars like our Sun and massive, early-type stars \citep{2023SSRv..219....7B}. They are considered the higher-mass counterparts of the classical T Tauri stars (CTTSs), representing a key transitional phase in stellar evolution. Typically aged between 2 and 8 million years, Herbig Ae/Be stars are intermediate-mass ($\sim$2–8 ,$M_\odot$) objects that continue to accrete material from their circumstellar disks. These stars remain surrounded by gas and dust, and exhibit a variety of observational signatures indicative of ongoing accretion and outflow activity and planet formation.
While the formation and accretion processes of low-mass stars are comparatively well understood, and the earliest stages of planet formation in their disks much investigated, the picture becomes less clear for higher-mass young stars. The magnetospheric accretion model, where the stellar magnetic field truncates the inner disk and channels infalling material along field lines, is well supported for CTTSs, but its applicability to Herbig Ae/Be stars remains uncertain. A key difference lies in the magnetic properties of these systems: whereas essentially all T Tauri stars possess strong magnetic fields \citep{2007ApJ...664..975J, 2022ApJ...925...21F}, only about 10\% of Herbig Ae/Be stars show detectable magnetic fields \citep{2018ApJ...852....5R}, and even these are substantially weaker. This low detection rate may reflect either genuinely weaker fields or more complex magnetic geometries that are difficult to measure with current techniques \citep{2023SSRv..219....7B}. Consequently, the mechanism by which accretion proceeds in Herbig Ae/Be stars likely differs from the well-established magnetically controlled regime seen in CTTSs. A further difference between CTTSs and Herbig Ae/Be stars is that planet formation around Herbig Ae/Be stars likely begins earlier and proceeds faster due to their more massive and warmer but shorter-lived disks \citep{Williams2011}.

This Mauve observing programme is designed to address these open questions through two complementary goals:
\begin{enumerate}
    \item To explore accretion physics in the intermediate-mass regime of Herbig Ae/Be stars, testing how the transition from magnetic to potentially non-magnetic accretion occurs.
    \item To determine whether Herbig Ae/Be stars exhibit periodic brightness variations analogous to the “dipper” and “burster” behaviour seen in CTTSs, and to study if this is linked to early planet formation.
\end{enumerate}

\subsubsection*{(i) Investigating accretion physics in Herbig Ae/Be stars}
\label{sec:Herbig}
\begin{figure}
    \centering
    \includegraphics[width=1\linewidth]{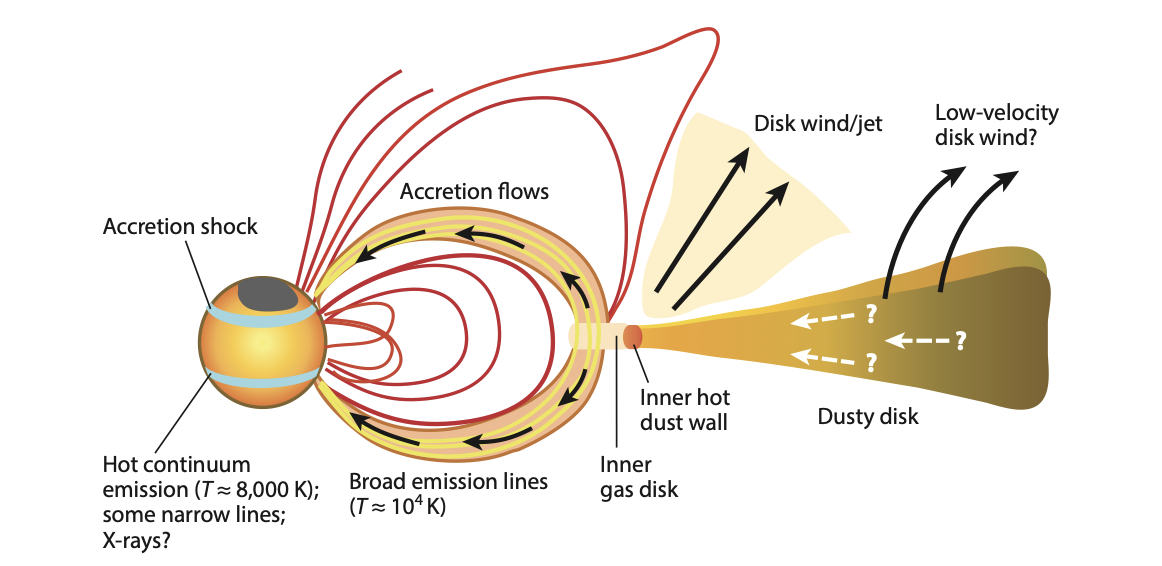}
    \caption{From \citet{2016ARA&A..54..135H}. The close circumstellar environment of a \ac{ctts}. Strong stellar fields truncate the accretion disk near the co-rotation radius, directing accreting material along the field lines where bright accretion shocks form.}
    \label{fig:krull1}
\end{figure}
\begin{figure}
    \centering
    \includegraphics[width=1\linewidth]{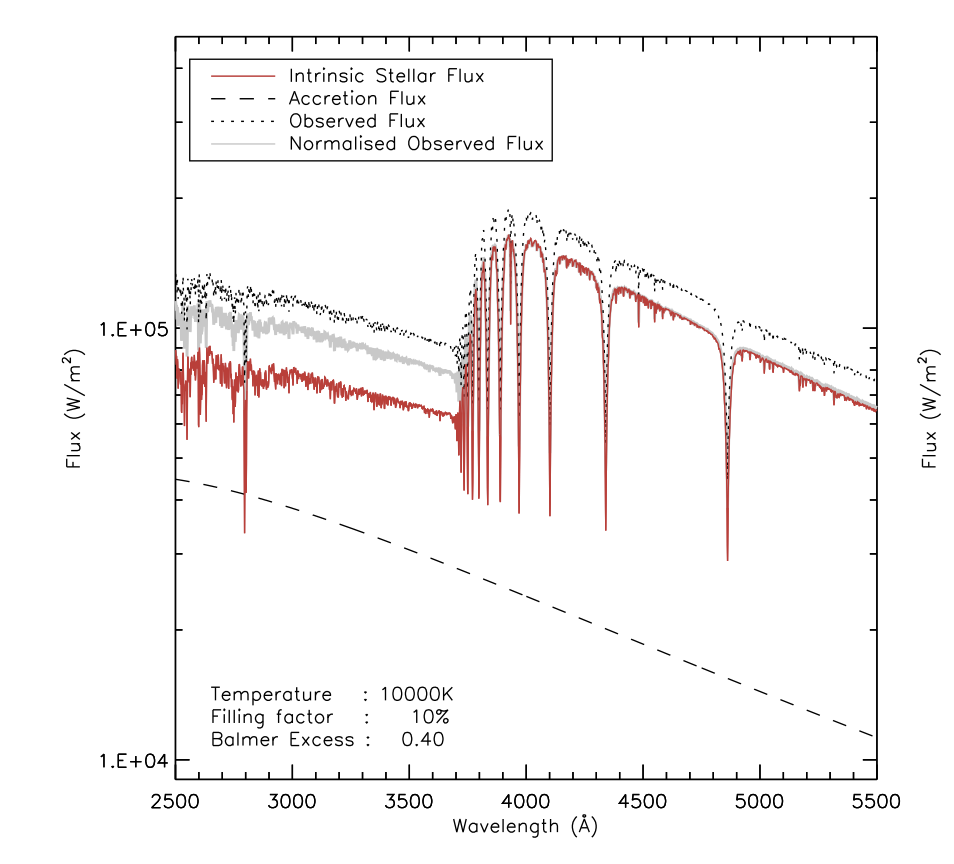}
    \caption{From \citet{2015MNRAS.453..976F}. The red spectrum is an A0 photosphere and the black dashed line shows the accretion flux. These sum to produce the observed black dotted spectrum, which when renormalised to match the photospheric (red) spectrum longward of the Balmer jump produces the light gray spectrum showing a substantial excess shortward of the Balmer jump.}
    \label{fig:krull2}
\end{figure}
\begin{figure}
    \centering
    \includegraphics[width=1\linewidth]{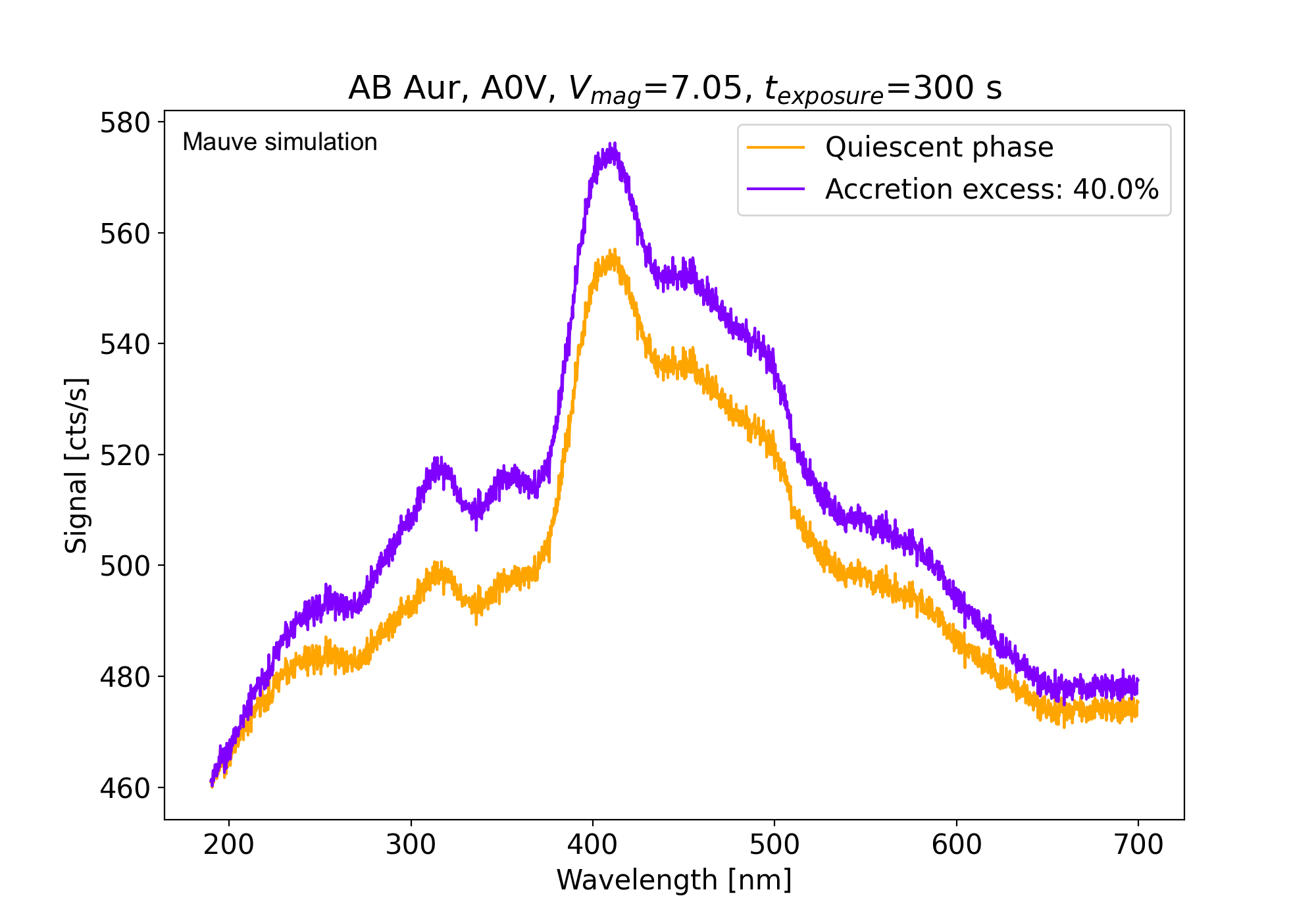}
    \caption{Simulated observed spectrum of AB Aur using MauveSim. Input spectrum is a star of the same spectral type (HD 143459) scaled to the V magnitude of AB Aur. Drawn in purple is how the accretion excess is expected to change the slope of the emission, specifically shortward of the Balmer jump \citep[e.g.][]{2015MNRAS.453..976F}.}
    \label{fig:krull3}
\end{figure}
\begin{figure}
    \centering
    \includegraphics[width=1\linewidth]{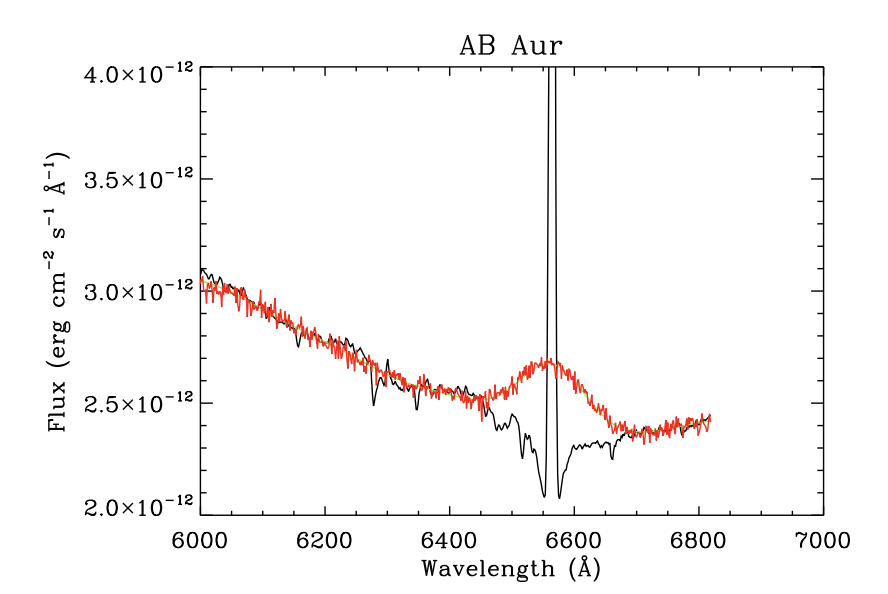}
    \caption{A spectrum of AB Aur in the neighbourhood of the H$\alpha$ line taken at a resolution of $\sim$1000 shown in black. This same spectrum convolved to the resolution of Mauve is shown in green and that spectrum observed at \ac{snr} = 100 is shown in red.}
    \label{fig:krull4}
\end{figure}

Monitoring the spectrophotometric modulation of light curves provides a powerful diagnostic of accretion in young stars, as variability often reflects changes in the accretion geometry or rate. Variability studies are therefore an excellent tool for probing accretion physics across stellar masses.
For CTTSs, it is well established that accretion from the circumstellar disk to the stellar surface is regulated by a strong magnetic field (Fig.~\ref{fig:krull1}), which truncates the disk at approximately five stellar radii above the surface \citep{1998ApJ...509..802C}. Material follows magnetic field lines onto the star, impacting in compact regions--accretion “footpoints”--that cover a small fraction of the stellar surface. If the stellar dipole is misaligned with the rotation axis, or if higher-order field components dominate \citep{2011AN....332.1027G}, the footpoints become localised, producing rotationally modulated photometric variability as they move in and out of view \citep{2020AJ....159..273R}.
However, if the magnetic field is weak or absent, as is thought to be the case for most Herbig Ae/Be stars, the situation changes drastically. Extrapolating the field geometry inward suggests that accretion could occur over a much larger area of the stellar surface, potentially forming an equatorial accretion ring in the limit of negligible magnetic confinement. In this scenario, the accretion emission would show weak or no rotational modulation, an important observational signature that Mauve is well equipped to detect.
The continuum emission from accretion shocks in CTTSs typically has a temperature of around 10,000 K \citep{1993AJ....106.2024V}, similar to the effective temperature of an A0 star. At first glance, this similarity might make accretion emission in Herbig Ae/Be stars difficult to detect. However, \cite{2015MNRAS.453..976F} demonstrated that the accretion excess remains clearly visible in the spectral region shortward of the Balmer jump, providing a reliable diagnostic for accretion even in these hotter stars (Fig.~\ref{fig:krull2}).
Mauve is particularly well suited to study this phenomenon across a large number of targets. Simulations using MauveSim \citep{10.1093/rasti/rzaf045} for the Herbig Ae star AB Aur (V = 7.05) show that in a 300-second exposure, the accretion excess shortward of the Balmer jump is readily detectable. In Fig.~\ref{fig:krull3} the expected photospheric emission of the A0 star (shown in orange in the simulation) contrasts clearly with the purple excess due to accretion. By monitoring the variability of this excess emission over time, we can directly probe in the accretion rate and geometry in Herbig Ae/Be stars.
In addition to the blue continuum, accretion also produces strong hydrogen recombination emission lines, particularly H$\alpha$ and other Balmer lines, as gas flows along magnetic channels or accretes directly from the disk. Although Mauve’s low spectral resolution will broaden these lines as shown in Fig.~\ref{fig:krull4}, flux variations in H$\alpha$ remain a sensitive and independent tracer of accretion variability. Combining line and continuum diagnostics will provide a comprehensive picture of the accretion processes in these intermediate-mass stars.

\subsubsection*{(ii) Do Herbig Ae/Be stars exhibit dipper and burster behaviour?}
\begin{figure}
    \centering
    \includegraphics[width=1\linewidth]{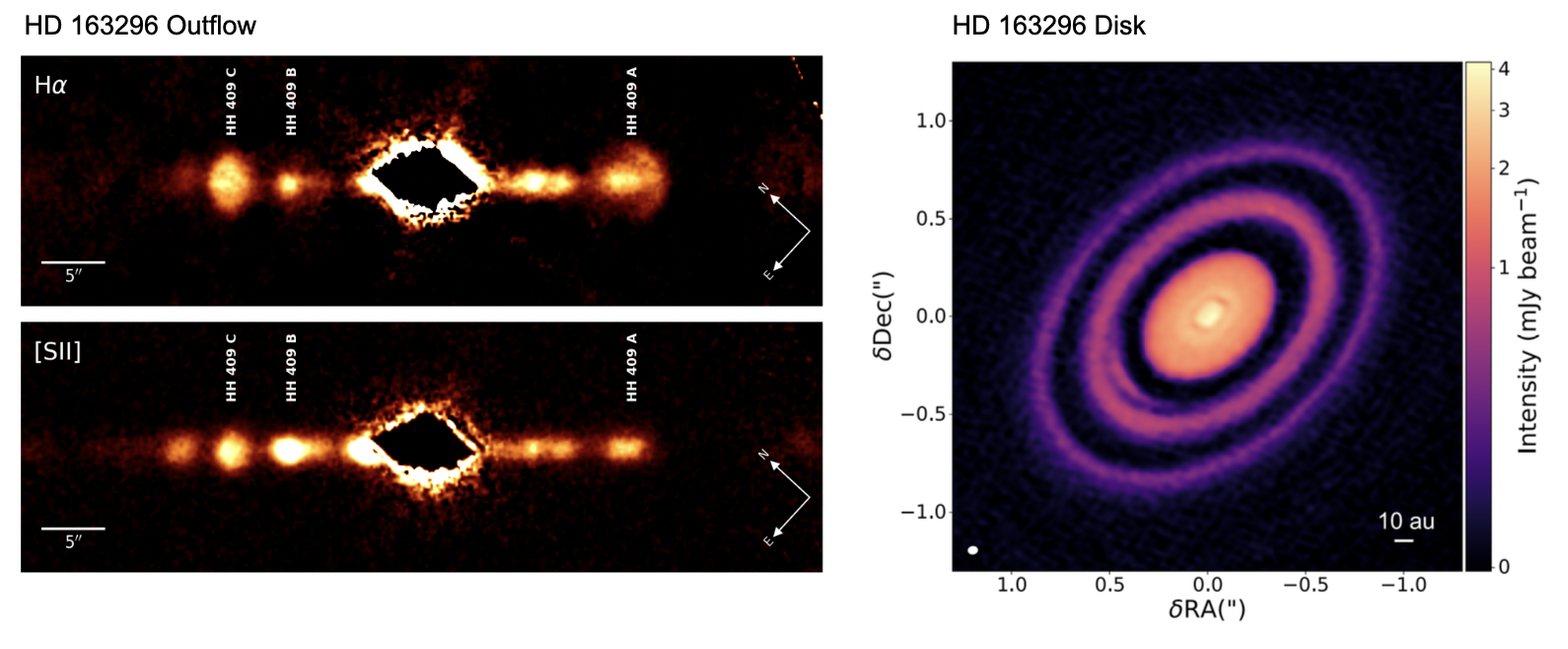}
    \caption{Left: VLT/MUSE observations of the bipolar jet driven by the Herbig Ae/Be star HD 163296 taken from \citet{2022A&A...663A..30K}. Right: Map of the HD 163296 disk recorded in the 1.25 mm continuum taken from \citet{Isella_2018}. The dust rings and sub-structures are interpreted as evidence that planet formation processes are active in the disk.}
    \label{fig:whelan1}
\end{figure}
\begin{figure}
    \centering
    \includegraphics[width=1\linewidth]{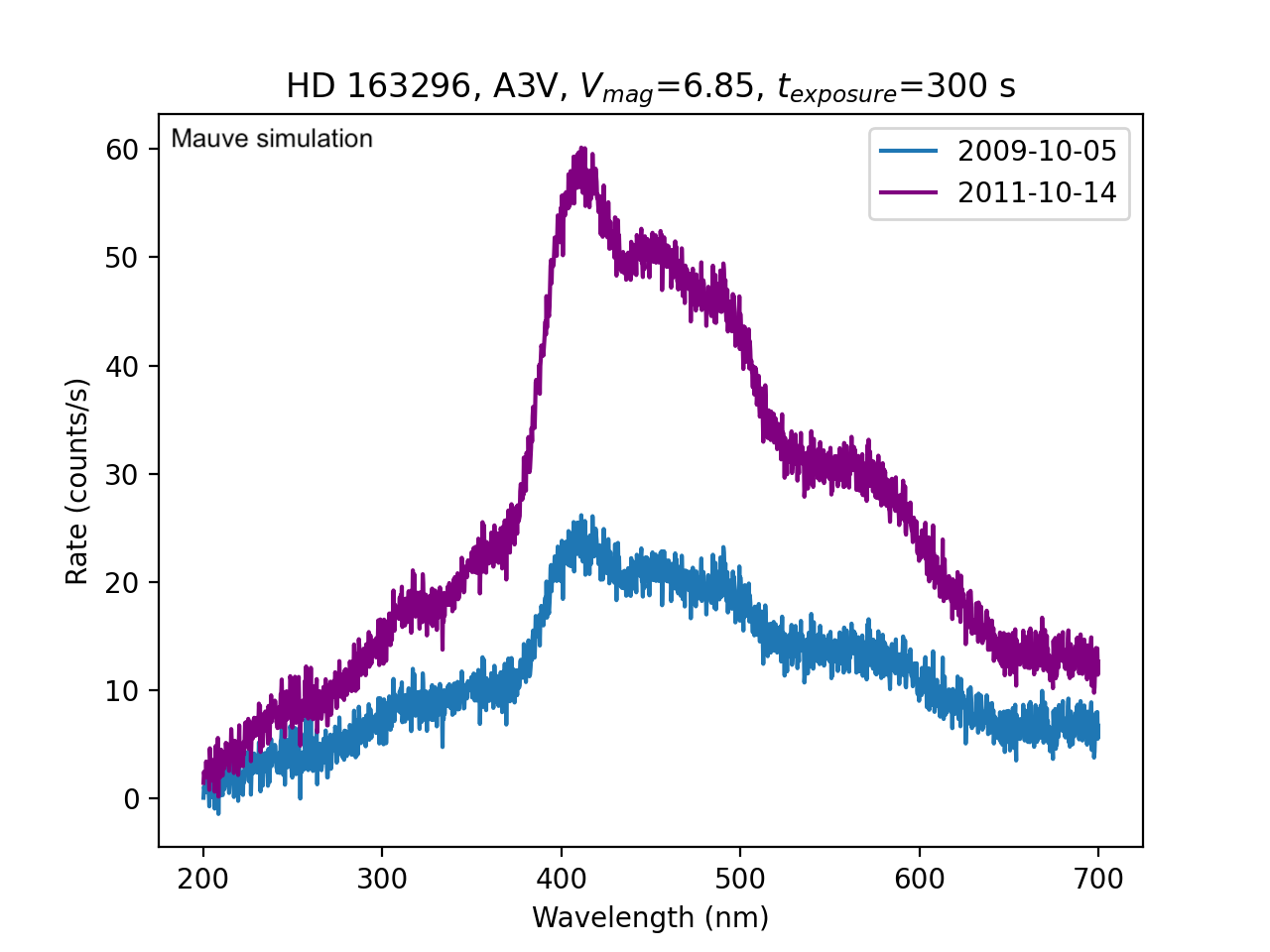}
    \caption{Simulated Mauve spectra of HD 163296, a Herbig Ae star with $V_{\rm mag} = 6.85$. The simulations are based on two epochs of X-Shooter observations that exhibit pronounced temporal variability. The noise levels in the simulated Mauve spectra correspond to an exposure time of 300 s.}
    \label{fig:whelan2}
\end{figure}

The relationship between star and planet formation is investigated by targeting several key properties in YSOs. These are accretion-ejection connection, variability, and accretion disk structure \citep{Pascucci2023, Fischer2023, Pinte2023}. Accretion-ejection connection and variability have primarily been observed at optical and \ac{ir} wavelengths \citep{2014A&A...565A..80W, Cody2017}, while sub-millimeter observations have revealed the structure of disks in amazing detail \citep{Andrews_2018}. The link between accretion and outflows is relevant to how planets form as outflows launched from the accretion disks are theorised to remove angular momentum thus driving accretion and disk evolution and shaping the initial conditions for planet formation \citep{Pascucci2025}. This activity is studied by targeting key emission lines in high-spectral and spatial resolution data allowing the mass accretion rate to be derived, and the kinematics, morphology, and physical conditions in the outflows to be mapped and accretion and outflow activity linked \citep{2022A&A...663A..30K, Birney2024}. Variability in the dynamical processes within the system, such as accretion onto the star, further impacts on properties of the accretion disks and thus the initial conditions for planet formation \citep{Das2025, Cieza2016}. Stellar variability can also point to activity in the inner disk associated with planet formation \citep{Kout2019, Petrov2015}. Finally, high angular resolution observations of the disks can reveal structure caused by orbiting planets or indeed in rare cases the protoplanets \citep{Andrews_2018, vanCapelleveen2025}.

Low mass YSOs are known to exhibit variability over a range of timescales \citep{Fischer2023} and dipper and burster behaviour has attracted much study \citep{Bonito_2023}. Dipper and burster behaviour refers to specific types of variability which last for days and are often linked to accretion and inner disk activity \citep{Cody2014}. Dippers experience periodic drops in brightness, which are thought to occur when material from their circumstellar disk intermittently blocks the star’s light \citep{Petrov2015}. This can be caused by the misalignment of the disk or the presence of large, dense structures within it, such as dust clumps or gas streams \citep{Empey2025}. Such structures occur during early planet formation. Bursters, on the other hand, exhibit sudden and dramatic increases in brightness, typically due to rapid, episodic accretion events where large amounts of material fall onto the star’s surface (also see (i) above). Although dipper and burster behaviour has most often been investigated in CTTSs, its implications for planet formation argue for systematic studies across a wider mass range.

Mauve will be used to construct light curves for a sample of Herbig Ae/Be stars to search for dipper and burster behaviour in the UV. As described above Herbig Ae/Be stars are strong accretors and they are also are associated with outflows, as seen in Figure~\ref{fig:whelan1} \citep{2022A&A...663A..30K}. They are known to be variable, exhibiting photometric and spectroscopic changes on timescales ranging from hours to months, and longer \citep{2011A&A...529A..34M} and they may have a lower incident of dipper behaviours \citep{Cody2025}. Also, the gas and dust in their circumstellar disks has been spatially resolved with unprecedented detail with ALMA, revealing Keplerian disks that show convincing evidence for the presence of forming planets \citep{2022A&A...658A.112S}. Therefore, they are ideal targets for investigating the connection between variability, and planet formation processes at higher masses. The focus with Mauve will mainly be on sources that drive outflows and/or exhibit signs of ongoing planet formation within their disks. Particularly interesting targets will then be examined in greater detail using existing and forthcoming ESO VLT UVES and MUSE data. Figure~\ref{fig:whelan2} presents simulated Mauve spectra based on two X-Shooter observations of the Herbig Ae star HD 163296 ($V_{\rm mag} = 6.85$), illustrating the variability between epochs. 
Mauve will monitor HD~163296 and similar sources with a typical cadence of 1-3 observations per day over timescales ranging from 40 to 80 days.

\subsection{Binaries in Exotic Stellar Populations}

A significant number of stars in the galaxy are formed in binaries or multiple systems, which give rise to complex interactions happening among the stars. Stars in such systems have thus different evolutionary trajectories compared to genuine single stars. These interactions lead to the formation of what are generally called exotic or non-canonical stellar populations, such as \acp{bss}, \acp{yss}, \acp{rss}, \acp{ssg}, subdwarfs (sdB, sdO), Li-rich stars, and so on. The exact origin of each of these types of stellar exotica is still wrapped in mystery. 

One of the intriguing products of stellar interactions are \acp{bss}, which appear brighter and bluer (hotter) than the normal turn-off stars, lying along an extension of the main sequence in the optical colour-magnitude diagram (CMD). In spite of the numerous photometric and spectroscopic studies on BSSs, their origin and evolution are still debated \citep{2015ASSL..413.....B}. To date, two leading formation channels suggested to account for their formation in clusters are: mass transfer in binaries \citep{1964MNRAS.128..147M} and stellar collisions or mergers \citep{1976ApL....17...87H}. Observational studies of \acp{bss} suggest that
a combination of all the formation channels is prevalent, but their importance varies based on many factors, including the stellar density of their host environments. Most \acp{bss} in low-density environments such as open clusters are found in binaries, with approximately 76\% in NGC\,188 with a range of orbital periods and eccentricities \citep{2009Natur.462.1032M}. \acp{bss} resulting from binary mass transfer usually have close companions, typically compact objects such as white dwarfs, unless the binary system was disrupted by supernova kicks. On the other hand, \acp{bss} formed through mergers, either from stellar collisions or binary mergers, typically do not have close companions. If these mergers occur within triple or higher-order multiple systems, the resulting \ac{bss} can have a companion \citep{1999ASPC..169..432I, 2009ApJ...697.1048P}. Several observational studies of \acp{bss} in the field and open clusters have detected white dwarf companions to them, implying that mass transfer may be a likely formation route \citep{2014ApJ...783L...8G, 2015ApJ...814..163G, 2019ApJ...885...45G, 2019ApJ...882...43S, 2023MNRAS.525.1311P}. \acp{bss} are thought to form via mass transfer occurring during different evolutionary phases, classified as case A, B, or C. These phases refer to the onset of mass transfer during the main sequence, before core helium burning, or after the exhaustion of core helium in the donor star \citep{1967ZA.....65..251K, 1970A&A.....7..150L}. 

Other class of \ac{uv}-bright stars are sub-dwarfs that are hot (O and B type) stars, mostly burning helium in their core (0.5 $M_\odot$) \citep{2009ARA&A..47..211H, 2016PASP..128h2001H}. They are hotter and more compact than canonical helium-burning stars, where a mix of stellar evolutionary processes, mass loss, binarity, and maybe mergers play a major role, which is demonstrated by the existence of several sub-classes among subdwarfs. A large amount of mass loss, required to form subdwarfs at the point of He burning, is difficult to explain in the context of single-star evolution and remains a missing piece of puzzle in stellar evolution theory \citep{2024A&A...686A..25D}. A large fraction of field sdB stars are found in close binaries with white dwarf or very low-mass main sequence companions, which must have gone through a common-envelope phase of evolution. 

\begin{figure*}
    \centering
    \includegraphics[width=0.48\linewidth]{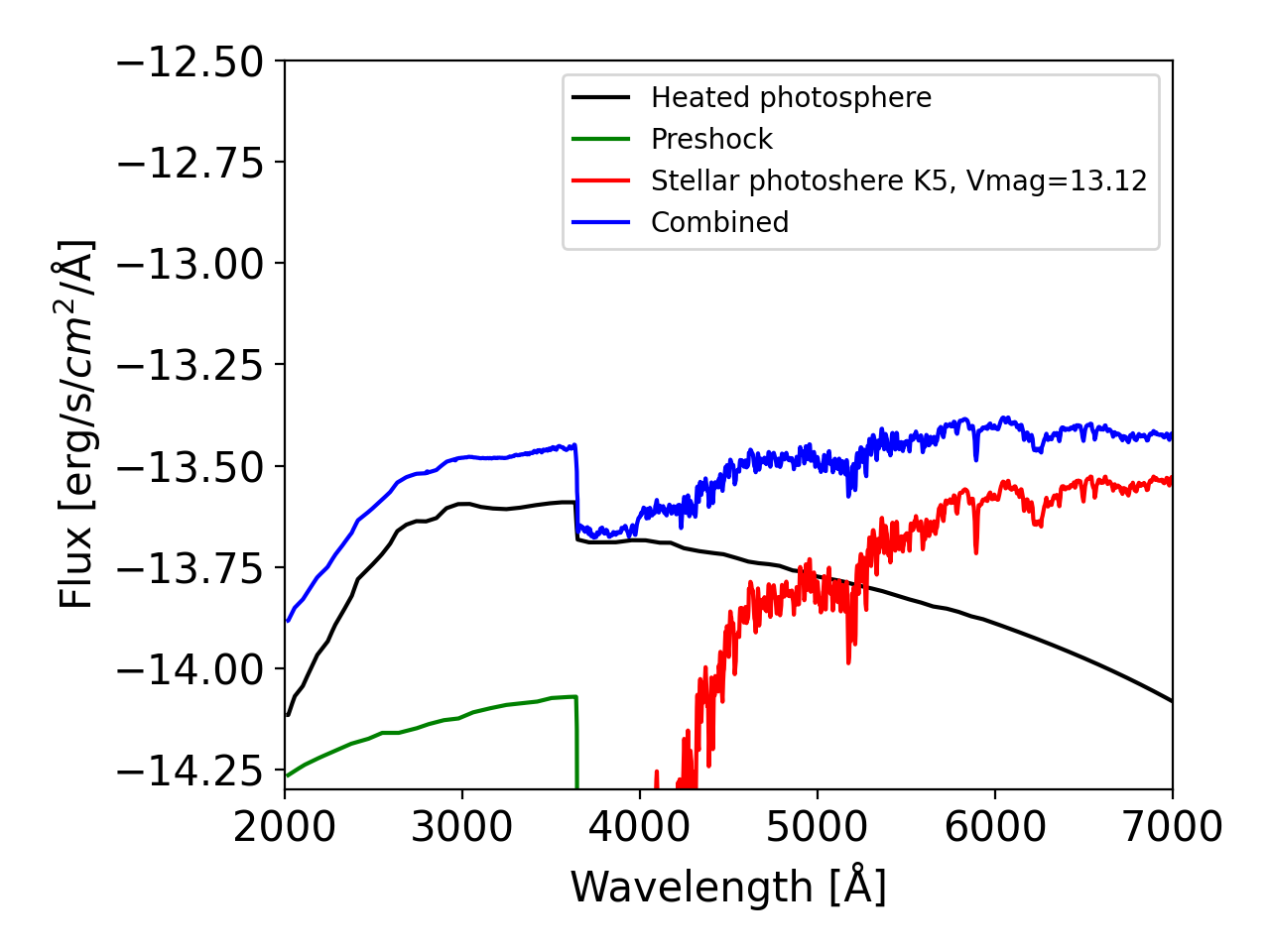}
    \hfill
    \includegraphics[width=0.48\linewidth]{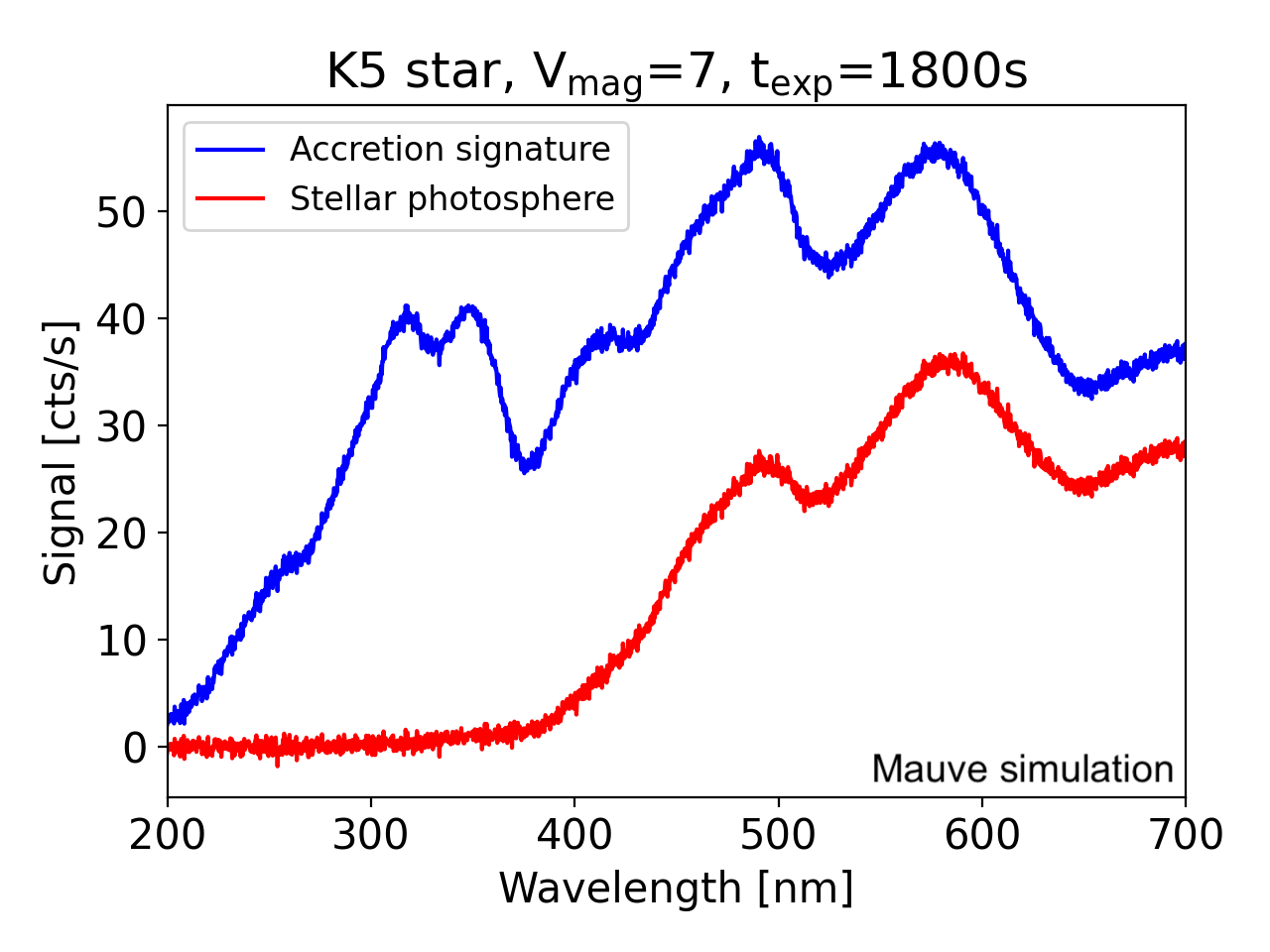}
    \caption{Left: SED of a pre-main sequence star at V$_{\rm mag}$=13.12 displaying accretion (blue line) which includes contributions from the preshock (green line), heated photosphere–postshock (black line) and stellar photosphere (red line). Figure reproduced from \citet{2016ARA&A..54..135H}. Right: Mauve simulated data for the accreting model versus the non-accreting (quiet) photosphere for a brighter K5 star.}
    \label{fig:ttauri}
\end{figure*}
Another anomalous class of exotic stellar populations that defy canonical single stellar evolution are \acp{ssg}, which lie on the redder side of the main sequence and are fainter than the normal subgiants in an optical CMD (see \citep{2018MNRAS.481..226S} and references therein). Most of the \acp{ssg} found in the field are in binaries and are chromospherically active; they are thought to be stars that are presently losing mass to an unseen
companion \citep{2017ApJ...840...66G}. Finally, Li-rich stars are a rare class of evolved stars that exhibit an abnormally high abundance of lithium in their atmospheres \citep{1982ApJ...255..577W}. They are observed at all evolutionary phases and their origin is still unclear. Among the various possible explanations in the literature, there are two different ideas involving a binary origin for the lithium enhancement, either mass transfer from an AGB star or a companion spinning up the visible star, thereby enhancing the lithium content on its surface \citep{1999ApJ...510..217S, 2019ApJ...880..125C}.

From the observational perspective, high-quality photometric and spectroscopic data are crucial for accurately determining the properties of exotic stars and discerning their formation mechanisms. Therefore, \ac{uv} imaging and spectroscopy offer a unique window to identify hot companions to the above-mentioned stars if present, thereby helping to constrain their formation and differences, if any, between the field and clusters populations. The \ac{nuv} spectra will complement the existing spectra in visible {wavelengths} to estimate stellar parameters ($L$, $T_{\rm eff}$, $R$) using multi-wavelength \acp{sed}, test evolutionary models, and derive the properties of binary components if present. The other advantage of observing in \ac{uv} is the large amplitude variations compared to optical and \ac{ir} regimes in the case of pulsating stars, and stars exhibiting any kind of stellar activity. \ac{uv} spectra will also enable the discernment of the evolutionary pathways of hot stars in the field and compare their properties with those in star clusters.

Three main pieces of evidence will be used in the search and characterisation of binaries; specifically:
\begin{enumerate}
    \item Any \ac{uv} excess which might indicate the presence of a hot companion. {Our candidates are mostly low-mass (<2 Msun) stars in various evolutionary phases, like main sequence, subgiant branch, red giant branch, hot subdwarfs. The hot companions are expected to be mostly white dwarfs or extremely low mass stars, with masses mostly in the range 0.1 to 0.5 Msun, thus we will be finding mostly unequal mass binaries.}
    \item Flux variability will be studied to determine if there are Keplerian motions (eclipses). {Eclipses will be easier to find for close binaries which, depending on the radius of the brighter star, will be on orbits that can range from one to several hundred days. However, shorter periods below about 10 days will be disproportionally more likely to produce eclipses.}
    \item Activity indicators will be used to disentangle stellar activity from Keplerian motions. {Activity can be a natural result of the stellar properties, such as for example for low mass main sequence stars, but can also be stimulated by close binary interactions. To distinguish between the two, we will use the whole target list, including low-mass stars from other themes in a collaborative effort, to establish the baseline of "normal" activity for different types of stars.}
\end{enumerate}
Mauve will be particularly useful in this respect for the first point, because it covers the \ac{nuv}
spectral region, for the second point because it will be possible to dedicate a sufficient amount
of time to monitor each star, and for the third point because of the wide range of wavelengths covered, including many of the typical activity indicators (such as Ca H\&K, NaD, H$\alpha$) and even some (like Mg II) which are not visible from the ground. While Mauve cannot resolve these lines due to its low resolution, light curve data across these wavelengths can still be used to detect variability. Moreover, when variability is detected, the comparison of the red and blue \ac{uv} variability will allow us to separate activity and pulsations (which change with passbands and thus colours) from Keplerian motions. 

The three observables indicated above will later be combined with existing literature information and/or follow-up observations, including spectral time series. For example, by combining spectral time series with Mauve light curves, we will be able to determine all orbital parameters, including the masses of the binary components, which are crucial to perform stellar evolution computations and reconstruct the paths of formation for these stars. Additionally, literature or follow-up information about the chemical composition of the star will provide further clues on the evolutionary stage of the companion during mass transfer, if any.

\section{Discussion \& Conclusions}
In this paper we have presented the science themes prioritised  by the Mauve Science Team for the first year of Mauve's operations. Initial simulations with MauveSim suggest strong performance across the different scientific goals and observational approaches presented here. However, the actual performance of Mauve will be confirmed through in-orbit testing and calibration, therefore, as the mission evolves, Mauve’s scientific focus and observational strategy will need to remain flexible.
New members are expected to join the survey collaboration throughout Year 1, with additional observational hours released periodically. Future science themes may focus on the detailed characterisation of accretion signatures in young, low-mass stars, as illustrated by the simulation in Fig.~\ref{fig:ttauri}.
Future publications will expand on Mauve's science operations, data analysis, satellite performances and science results, {including reporting on payload and detector ageing throughout the mission lifetime.}

Further into the future, Mauve$^{+}$ is part of a BSSL roadmap to deliver more performing UV satellites for astronomy compared to Mauve. Mauve$^{+}$ will have a larger telescope and higher spectral resolving power to resolve single spectral lines. Mauve$^{+}$'s preliminary design includes a 25+ cm telescope coupled with a spectrograph (\textit{R} $\approx$ 1000) covering 180-500 nm. These characteristics will allow to use this space telescope to investigate the high-energy environments of exoplanet host stars, monitor accretion processes and stellar flaring and provide critical UV coverage for transient multi-messenger events, with the capability of resolving individual spectral lines.

\section{Authors Contributions}

The Mauve Science Team has prepared this group publication to show the range of interests of the survey collaboration. While the survey programme is a collaborative initiative, and members are involved in multiple science themes, each member contributed to the publication as follows: 

\begin{itemize}
    \item Radiation mechanisms of stellar flares on M-dwarfs and young Sun-like stars: Hiroyuki Maehara and Kosuke Namekata. 
    \item Probing coronal mass ejections through UV dimming signatures: Chuanfei Dong and Hongpeng Lu. 
    \item Quiescent UV Emission in Low-Mass Stars: Marcel A. Agüeros, Alejandro Núñez and Krisztián Vida.
    \item Young Exoplanet Hosts and HWO: Girish M. Duvvuri and Keivan G. Stassun. 
    \item  The Classical Be star survey: T. A. A Sigut and Anusha Ravikumar.
    \item Accretion variability of Herbig Ae/Be stars and implications for planet formation: Christopher M. Johns–Krull and Don Dixon. 
    \item  Do Herbig Ae/Be stars exhibit dipper and burster behaviour: Emma T. Whelan, Jamie J. Stewart and Patrick F. Flanagan. 
    \item Binaries in Exotic Stellar Populations: Elena Pancino and Sharmila Rani. 
\end{itemize}

\section*{Acknowledgements}
The Mauve project has received funding from the European Union’s Horizon Europe research and innovation programme under grant agreement No. 101082738.
T.\ A.\ A.\ Sigut is grateful for support from the Fredrick Hunt Physics and Astronomy Fund and the Natural Sciences and Engineering Research Council of Canada (NSERC) through the Discovery Grant program. C. Dong acknowledges support from NASA grant No. 80NSSC23K1115, the Alfred P. Sloan Research Fellowship, and the IBM Einstein Fellow Fund at the Institute for Advanced Study, Princeton. E. Pancino and S. Rani are co-funded by the European Union (ERC-2022-AdG, "StarDance: the non-canonical evolution of stars in clusters", Grant Agreement 
101093572, PI: E. Pancino). K. Vida's research is funded by the Hungarian National Research, Development and Innovation Office grant \'Elvonal KKP-143986.

\section*{Data Availability}
All data for this paper are contained within the article; BSSL is able to share simulated data upon request.
 

\printacronyms[name=ACRONYMS, template=tabular]



\bibliographystyle{rasti}
\bibliography{example} 








\bsp	
\label{lastpage}
\end{document}